\documentclass[aps,prd,twocolumn,showpacs,10pt,superscriptaddress,preprintnumbers,nofootinbib]{revtex4-1}
\usepackage[utf8]{inputenc}
\usepackage{amsmath,amssymb,bm,slashed,braket}
\usepackage{graphicx}
\usepackage{epstopdf}
\usepackage{float}
\usepackage[dvipsnames,table]{xcolor}
\usepackage[normalem]{ulem}
\usepackage{colortbl}
\usepackage{multirow, booktabs}
\usepackage{makecell}
\usepackage{hhline}
\usepackage{dsfont}
\usepackage{verbatim}
\usepackage[colorlinks=true,
            linkcolor=blue,
            urlcolor=blue,
            citecolor=purple,          
            bookmarks=true,
            bookmarksnumbered=true,
            breaklinks=true,
            pdfpagemode=Fullscreen,
            pdfstartview=FitBH]{hyperref}

\allowdisplaybreaks[4]
\usepackage[capitalise]{cleveref}
\usepackage{stackengine}
\usepackage{orcidlink}
\usepackage{nicematrix}
\usepackage{ctable}
\usepackage{pifont}
\usepackage{cancel}
\usepackage{tikz} 
\usepackage{tkz-euclide}
\usetikzlibrary{backgrounds} 
\usetikzlibrary{decorations.pathmorphing}
\usetikzlibrary{arrows.meta}
\tikzset{
mystyle/.style={line width=1, baseline, scale=0.6, every node/.style={scale=1}},
v/.style={decorate, draw, decoration={snake, segment length=2.mm, amplitude=0.5mm}},
f/.style={draw, decoration={markings,mark=at position #1 with {\arrow[]{Latex[length=1.5mm,width=1.5mm]}}},
    postaction={decorate},node contents=#1},
f/.default=.6,
fb/.style={draw,decoration={markings,mark=at position #1 with {\arrowreversed[]{Latex[length=1.5mm,width=1.5mm]}}},
    postaction={decorate},node contents=#1},
fb/.default=.6,
s/.style={dashed,draw, decoration={markings,mark=at position #1 with {\arrow[]{Latex[length=1.5mm,width=1.5mm]}}},
    postaction={decorate},node contents=#1},
s/.default=.6,    
sb/.style={dashed,draw,decoration={markings,mark=at position #1 with {\arrowreversed[]{Latex[length=1.5mm,width=1.5mm]}}},
    postaction={decorate},node contents=#1},
sb/.default=.4,
snar/.style={dashed,draw,line width =1.25pt},
cross/.style={cross out, draw=black, minimum size=2*(#1-\pgflinewidth), inner sep=0pt, outer sep=0pt}, 
         }

\newcommand{\C}{{\tt C}}

\newcommand{\calO}{{\cal O}}

\newcommand{\hc}{\text{h.c.}}

\newcommand{\tL}{{\tt L}}
\newcommand{\tR}{{\tt R}}

\begin{document}

\title{Chiral perturbation theory for baryon-number-violating nucleon decay into a vector meson}

\author{Yi Liao\,\orcidlink{0000-0002-1009-5483}}
\email{liaoy@m.scnu.edu.cn}
\author{Xiao-Dong Ma\,\orcidlink{0000-0001-7207-7793}}
\email{maxid@scnu.edu.cn}
\author{Hao-Lin Wang\,\orcidlink{0000-0002-2803-5657}}
\email{whaolin@m.scnu.edu.cn}
\affiliation{State Key Laboratory of Nuclear Physics and
Technology, Institute of Quantum Matter, South China Normal
University, Guangzhou 510006, China}
\affiliation{Guangdong Basic Research Center of Excellence for
Structure and Fundamental Interactions of Matter, Guangdong
Provincial Key Laboratory of Nuclear Science, Guangzhou
510006, China}

\begin{abstract}
In a recent work [New chiral structures for baryon number violating nucleon decays, \href{https://arxiv.org/abs/2504.14855}{arXiv:2504.14855}.] we identified generic baryon-number-violating (BNV) structures containing triple light quarks and achieved their leading-order chiral realizations involving octet pseudoscalars and baryons. Although many two-body nucleon decays into a vector meson have been experimentally searched for and stringently constrained, a consistent theoretical framework for their calculation is still lacking. In this Letter, we fill the gap by implementing chiral matching of all these triple-quark interactions onto hadronic interactions involving octet vector mesons, baryons, and pseudoscalars. This paves the way for a consistent and comprehensive study of all relevant BNV processes. As an illustration of application, we show how degeneracy in the parameter space of Wilson coefficients can be broken by synthesizing experimental constraints on nucleon decays into a vector or pseudoscalar meson when relevant hadronic low-energy constants can be reasonably determined. 

\end{abstract}
\maketitle

\vspace{0.1cm}{\bf Introduction.~}
Baryon-number-violating (BNV) interactions are of fundamental importance in particle physics and cosmology due to their connections to major unresolved questions, such as the origin of 
matter-antimatter asymmetry in the Universe \cite{Sakharov:1967dj}, the 
nature of dark matter, and the mechanism
for tiny neutrino mass. 
Investigating these interactions can provide critical insights into physics beyond the standard model and guide the search for new particles and forces. 

Nucleon decay that violates baryon number by one unit ($\Delta B=1$) serves as a golden channel for probing BNV interactions, owing to the vast number of nucleons available in terrestrial detectors. The two-body nucleon $\texttt{N}$ (either a proton $p$ or a neutron $n$) decays to an octet pseudoscalar meson $M$ and a lepton $l$ (either a charged lepton $\ell$ or a neutrino $\nu$)
have been extensively searched for by large-fiducial-mass experiments, including  IMB \cite{Irvine-Michigan-Brookhaven:1983iap}, 
SNO+ \cite{SNO:2018ydj}, 
KamLAND \cite{KamLAND:2015pvi},
Kamiokande \cite{Hirata:1988ad}, 
and Super-Kamiokande \cite{Takhistov:2016eqm}.
Although no positive signals have been observed, these experiments have established stringent limits on the occurrence of such decays. 
Besides the pseudoscalar modes, significant experimental endeavors have been undertaken in nucleon 
decays into an octet vector meson 
\cite{Super-Kamiokande:2017gev,McGrew:1999nd,Kamiokande-II:1989avz,Frejus:1990myz,Super-Kamiokande:2012ngt}. 
Their experimental limits are comparably stringent to pseudoscalar channels; see the second column of \cref{tab:N2lXbound}.
The next generation of neutrino experiments, such as DUNE \cite{DUNE:2020ypp}, Hyper-Kamiokande \cite{Hyper-Kamiokande:2018ofw}, JUNO \cite{JUNO:2015zny}, and THEIA \cite{Theia:2019non}, are expected to further improve sensitivity to these and other nucleon decay modes. 

Theoretically, effective field theory (EFT) has emerged as an increasingly vital tool for investigating nucleon decay in a model-independent manner. 
In a recent work \cite{Liao:2025vlj}, we have examined the most general $\Delta B=1$ interactions in the low-energy effective field theory (LEFT)  that involve triple fields of  light quarks. 
We found that they are classified into four irreducible representations (irreps)
under the QCD chiral group 
$G_\chi\equiv\rm SU(3)_{\tt L}\times SU(3)_{\tt R}$: 
$\pmb{8}_{\tt L}\otimes \pmb{1}_{\tt R}$, 
$\bar{\pmb{3}}_{\tt L}\otimes \pmb{3}_{\tt R}$,
$\pmb{6}_{\tt L}\otimes \pmb{3}_{\tt R}$,
$\pmb{10}_{\tt L}\otimes \pmb{1}_{\tt R}$, plus their chirality partners under the interchange of chiralities ${\tt L}\leftrightarrow{\tt R}$. 
Restricting to dimension-6 (dim-6) operators in LEFT, only the first two irreps (and their chirality partners) are possible, and correspond to the ones known since the 1980s \cite{Wilczek:1979hc,Ellis:1979hy,Weinberg:1979sa,Weinberg:1980bf,Abbott:1980zj,Beneito:2023xbk,Broussard:2025opd}, while the other two are new and can only be realized by higher dimensional operators. We have also achieved in that work the leading-order (LO)  chiral realizations for all of the above irreps that involve the nucleon and octet pseudoscalar mesons. Again, while the chiral realizations for the dim-6 operators reproduce the known result in~\cite{Claudson:1981gh}, the chiral realizations of the operators in the two new irreps are new and turn out to be nontrivial.

\begin{figure}[t]
\centering
\begin{tikzpicture}[scale=0.8,every node/.style={scale=0.8}]
\begin{scope}[shift={(1,1)}] 
\filldraw[draw=gray!20!white,fill=gray!20!white] (-2.1,1.) rectangle (4.1,4.5);
\node at (1,3){${\calO}_{\nu uds}^{\tL\tR,x},\,{\calO}_{\nu dsu}^{\tL\tR,x},\,{\calO}_{\nu sud}^{\tL\tR,x} $ };
\node at (1,2.5){$\,{\calO}_{\nu dsu}^{\tL\tL,x},\,{\calO}_{\nu sud}^{\tL\tL,x} $ };
\draw[f,thick,purple] (0,3.367) -- (-1,4);
\node at (-1,4.2){$p\to\bar\nu_x K^+$ };
\draw[f,thick,purple] (2,3.367) -- (3,4);
\node at (3,4.2){$n\to\bar\nu_x K^0$ };
\draw[f,thick,orange] (0,2.133) -- (-1,1.5);
\node at (-1,1.3){$p\to\bar\nu_x K^{*+}$ };
\draw[f,thick,orange] (2,2.133) -- (3,1.5);
\node at (3,1.3){$n\to\bar\nu_x K^{*0}$ };
\draw[thick,blue!80] (1,2.75) ellipse (1.6 and 0.8);
\end{scope}
\end{tikzpicture}
\vspace{-0.5em}
\caption{An example of complementarity between pseudoscalar and vector meson modes of nucleon decay in constraining dim-6 BNV interactions involving a strange quark.}
\label{fig:p2vKstar}
\end{figure}
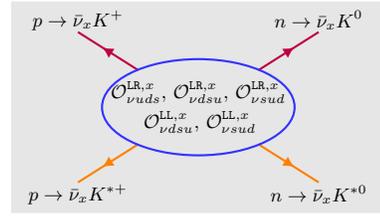
 
With the above experimental advances we think it is timely to develop a chiral perturbation theory for octet vector mesons and baryons. As we will analyze later, such a consistent theory is not available in the literature. As a bonus, this will facilitate a complementary exploration of BNV interactions through both pseudoscalar and vector meson channels. An example is shown in \cref{fig:p2vKstar} for the dim-6 operators containing a strange quark that can be probed by two-body nucleon decays into either a pseudoscalar or a vector kaon.

\vspace{0.1cm}{\bf General BNV triple quark interactions.} 
To set the stage, we recall our recent results on BNV interactions involving triple quarks~\cite{Liao:2025vlj}.
In LEFT or its extension by a new light field, a general BNV operator involving triple light quarks without being acted upon by a derivative always fall into one of the following four independent structures and their chirality partners, 
\begin{align}
{\cal O}_{a}^{yzw} & = \overline{\Psi_{a}}{\cal N}_{yzw}^{\tL\tL},
&
{\cal O}_{b}^{yzw} & =\overline{\Psi_{b}}{\cal N}_{yzw}^{\tR\tL}, 
\nonumber\\
{\cal O}_{c}^{yzw} & = \overline{\Psi_{c,\mu}}
{\cal N}_{yzw}^{\tL\tR,\mu},
&
{\cal O}_{d}^{yzw} & = \overline{\Psi_{d,\mu\nu}}
{\cal N}_{yzw}^{\tL\tL,\mu\nu}, 
\label{eq:LEFT3qO}
\end{align} 
where the $\Psi$s are nonquark fields, and the ${\cal N}$s are formed with three chiral quark fields in a definite irrep of
$G_\chi$, 
\begin{subequations}
\label{eq:Nyzw}
\begin{align}
{\cal N}_{yzw}^{\tL\tL} & =  q_{\tL, y}^\alpha (\overline{ q_{\tL, z}^{\beta \C} } q_{\tL, w}^\gamma )\epsilon_{\alpha \beta \gamma}\in \pmb{8}_\tL \otimes  \pmb{1}_\tR , 
\\
{\cal N}_{yzw}^{\tR\tL} & = q_{\tR, y}^\alpha (\overline{ q_{\tL, z}^{\beta \C} } q_{\tL,w}^\gamma)\epsilon_{\alpha \beta \gamma} \in 
\bar{\pmb{3}}_\tL \otimes \pmb{3}_\tR , 
\\
{\cal N}_{yzw}^{\tL\tR,\mu} & = q_{\tL,\{y}^\alpha (\overline{ q_{\tL, z\}}^{\beta \C} } \gamma^\mu q_{\tR,w}^\gamma)\epsilon_{\alpha \beta \gamma}
\in \pmb{6}_\tL \otimes \pmb{3}_\tR,
\\
{\cal N}_{yzw}^{\tL\tL,\mu\nu} & =  q_{\tL, \{y}^\alpha (\overline{ q_{\tL, z}^{\beta \C} } \sigma^{\mu\nu} q_{\tL, w\} }^\gamma )\epsilon_{\alpha \beta \gamma}\in \pmb{10}_\tL \otimes  \pmb{1}_\tR. 
\end{align}
\end{subequations}
Here $\alpha,~\beta,~\gamma$ denote color and $y,~z,~w=1,~2,~3$ denote three light quarks with $q_{1,2,3}=u,~d,~s$. 
The curly brackets indicate symmetrization in flavor indices of like-chirality fields. 
When $\Psi$s are a single fermion field, i.e., when the ${\cal O}$ operators are of dim-6, $\calO_{a,b}$ recover the two known operators while $\calO_{c,d}$ vanish identically.

Following our previous work \cite{Fan:2024gzc}, we 
organize in matrix form the ${\cal N}$s in the irreps ${\bf 8}_\tL\otimes {\bf 1}_\tR$ and $\bar{\pmb{3}}_\tL \otimes \pmb{3}_\tR$:
\begin{subequations}
\label{eq:3qpart}
\begin{align}
{\cal N}_{{\bf 8}_\tL\otimes {\bf 1}_\tR}
& =
\begin{pmatrix}
{\cal N}^{\tL\tL}_{uds}  &  {\cal N}^{\tL\tL}_{usu}  & {\cal N}^{\tL\tL}_{uud}  
\\[1pt]
{\cal N}^{\tL\tL}_{dds}  & {\cal N}^{\tL\tL}_{dsu} & {\cal N}^{\tL\tL}_{dud}  
\\[1pt]
{\cal N}^{\tL\tL}_{sds} & {\cal N}^{\tL\tL}_{ssu} & {\cal N}^{\tL\tL}_{sud}
\end{pmatrix},
\\
{\cal N}_{\bar{\pmb{3}}_\tL \otimes \pmb{3}_\tR } & = 
 \begin{pmatrix}
{\cal N}_{uds}^{\tR\tL}  
& {\cal N}_{usu}^{\tR\tL} 
& {\cal N}_{uud}^{\tR\tL} 
\\[1pt]
{\cal N}_{dds}^{\tR\tL}  
& {\cal N}_{dsu}^{\tR\tL} 
& {\cal N}_{dud}^{\tR\tL} 
\\[1pt] 
{\cal N}_{sds}^{\tR\tL}  
& {\cal N}_{ssu}^{\tR\tL} 
& {\cal N}_{sud}^{\tR\tL}
 \end{pmatrix}.
\end{align}
\end{subequations}
Together with ${\cal N}_{yzw}^{\tL\tR,\mu}$ and ${\cal N}_{yzw}^{\tL\tL,\mu\nu}$ in \cref{eq:Nyzw}, 
they transform under $(\hat L,\hat R) \in G_\chi$ as
\begin{subequations}
\begin{align}
{\cal N}_{\pmb{8}_\tL \otimes \pmb{1}_\tR} &\rightarrow
\hat L {\cal N}_{\pmb{8}_\tL \otimes \pmb{1}_\tR} \hat L^\dagger, 
\\%
{\cal N}_{\bar{\pmb{3}}_\tL \otimes \pmb{3}_\tR } &\rightarrow
\hat R {\cal N}_{  \bar{\pmb{3}}_\tL  \otimes \pmb{3}_\tR} \hat L^\dagger,
\\%
{\cal N}_{yzw}^{\tL\tR,\mu} 
&\rightarrow
\hat L_{yy'} \hat L_{zz'} \hat R_{w w'}  
{\cal N}_{y'z'w'}^{\tL\tR,\mu},
\\%
{\cal N}_{yzw}^{\tL\tL,\mu\nu} &\rightarrow
\hat L_{y y'} \hat L_{z z'} \hat L_{w w'}
{\cal N}_{y'z'w'}^{\tL\tL,\mu\nu} . 
\end{align} 
\end{subequations} 
The chirality partners of the ${\cal N}$s along with their chiral transformations are obtained from the above by swapping $\tL$ with $\tR$ and $\hat L$ with $\hat R$. 

In the Lagrangian of LEFT each operator ${\cal O}_i^{yzw}$ ($i=a,b,c,d$) is multiplied by a Wilson coefficient (WC) $C_i^{yzw}$ which retains the same flavor symmetry as the operator. We denote the product of $C_i^{yzw}$ and $\overline{\Psi_i}$ by the spurion field ${\cal P}_{yzw}^i=C_i^{yzw}\overline{\Psi_i}$. To facilitate chiral matching we assign chiral transformations to the spurion fields that are conjugate to those of ${\cal O}_i^{yzw}$, so that the effective interactions are apparently chiral invariant:
\begin{subequations}
\label{eq:chitran}
\begin{align}
{\cal P}_{\pmb{8}_\tL \otimes \pmb{1}_\tR}
& \to
\hat L {\cal P}_{\pmb{8}_\tL \otimes \pmb{1}_\tR} \hat L^\dagger,
\\%
{\cal P}_{\pmb{3}_\tL \otimes \bar{\pmb{3}}_\tR } 
& \to  
\hat L {\cal P}_{ \pmb{3}_\tL \otimes \bar{\pmb{3}}_\tR} \hat R^\dagger,
\\%
{\cal P}_{yzw}^{\tL\tR,\mu}
& \to
\hat L^*_{yy'} \hat L^*_{zz'} \hat R_{ww'}^*  
{\cal P}_{y'z'w'}^{\tL\tR,\mu},
\\%
{\cal P}_{yzw}^{\tL\tL,\mu\nu}
& \to \hat L_{yy'}^* \hat L_{zz'}^* \hat L_{ww'}^*
{\cal P}_{y'z'w'}^{\tL\tL,\mu\nu},
\end{align}
\end{subequations}
where ${\cal P}_{\pmb{8}_\tL \otimes \pmb{1}_\tR}$ and ${\cal P}_{\pmb{3}_\tL \otimes \bar{\pmb{3}}_\tR }$ are organized in matrix form, similar to \cref{eq:3qpart} but with ${\cal N}$ replaced by ${\cal P}$ throughout.
In \cref{eq:spu} of the Supplemental Material (SM), we present the explicit forms of these spurion fields related to the dim-6 LEFT BNV interactions. 
To summarize, the general $\Delta B=1$ Lagrangian involving triple light quarks can be compactly expressed as~\cite{Liao:2025vlj} 
\begin{align}
{\cal L}_{q^3}^{\slashed{B}} & = 
{\rm Tr} \big[  
  {\cal P}_{\pmb{8}_\tL \otimes \pmb{1}_\tR }
  {\cal N}_{\pmb{8}_\tL \otimes \pmb{1}_\tR } 
+ {\cal P}_{ \pmb{1}_\tL \otimes \pmb{8}_\tR }  
  {\cal N}_{  \pmb{1}_\tL \otimes \pmb{8}_\tR }
  \big]  
\nonumber 
\\
& + {\rm Tr} \big[ 
  {\cal P}_{\pmb{3}_\tL \otimes \bar{\pmb{3}}_\tR }
  {\cal N}_{\bar{\pmb{3}}_\tL \otimes \pmb{3}_\tR } 
+ {\cal P}_{\bar{\pmb{3}}_\tL \otimes \pmb{3}_\tR }
  {\cal N}_{\pmb{3}_\tL \otimes \bar{\pmb{3}}_\tR }
 \big] 
\nonumber 
\\
& + \big[
{\cal P}_{yzw}^{\tL\tR,\mu}
{\cal N}_{yzw,\mu}^{\tL\tR}
+ {\cal P}_{yzw}^{\tR\tL,\mu}
{\cal N}_{yzw,\mu}^{\tR\tL}
\big]
\nonumber 
\\%
& + \big[
{\cal P}_{yzw}^{\tL\tL,\mu\nu}
{\cal N}_{yzw,\mu\nu}^{\tL\tL}
+ 
{\cal P}_{yzw}^{\tR\tR,\mu\nu}
{\cal N}_{yzw,\mu\nu}^{\tR\tR}
\big]
 +\rm H.c.,
\label{eq:q3LEFT}
\end{align}
where summation over indices $y,z,w$ is implied.

\vspace{0.1cm}{\bf Chiral matching.~}
To obtain the hadronic counterparts of the quark-level interactions in \cref{eq:q3LEFT}, a consistent approach is to use chiral perturbation theory
(ChPT), which systematically accounts for interactions among the octet pseudoscalars, vectors, and baryons. 
There are several approaches to incorporating the vector mesons: 
the matter field approach \cite{Weinberg:1968de,Ecker:1989yg}, the hidden local symmetry \cite{Bando:1984pw,Fujiwara:1984mp}, the tensor field approach \cite{Gasser:1983yg,Ecker:1988te,Ecker:1989yg}, and the massive Yang-Mills method \cite{Gasiorowicz:1969kn,Kaymakcalan:1983qq,Gomm:1984at,Meissner:1987ge}. It has been demonstrated in \cite{Ecker:1989yg,Birse:1996hd} that all these approaches are equivalent. 
In this work, we adopt the matter field approach 
in terms of four-vector potentials that transform homogeneously under the nonlinear realization of the chiral symmetry. 

We arrange the octet meson and baryon fields in the matrix form,
\begin{subequations}
\begin{align}
\Sigma(x) & = \xi^2(x) = \exp\Big(\frac{i\sqrt{2}\Pi(x)}{F_0}\Big) ,  
\\
\Pi(x) & =   
\begin{pmatrix}
\frac{\pi^0}{\sqrt{2}}+\frac{\eta}{\sqrt{6}} & \pi^+ & K^+\\
\pi^- & -\frac{\pi^0}{\sqrt{2}}+\frac{\eta}{\sqrt{6}} & K^0\\
K^- & \bar{K}^0 & -\sqrt{\frac{2}{3}}\eta
\end{pmatrix},
\\
B(x) &=
\begin{pmatrix}
\frac{\Sigma^{0}}{\sqrt{2}}+\frac{\Lambda^0}{\sqrt{6}}  & \Sigma^+ & p \\
\Sigma^- & -\frac{\Sigma^{0}}{\sqrt{2}}+\frac{\Lambda^0}{\sqrt{6}} &  n \\ 
\Xi^- & \Xi^0 & - \sqrt{\frac{2}{3}}\Lambda^0
\end{pmatrix},
\\
V_\mu(x) &=
\begin{pmatrix}
\frac{\rho_\mu^{0}}{\sqrt{2}}+\frac{\phi_\mu^{(8)}}{\sqrt{6}} & \rho_\mu^+ & K_\mu^{*+} \\
\rho_\mu^- & -\frac{\rho_\mu^{0}}{\sqrt{2}}+\frac{\phi_\mu^{(8)}}{\sqrt{6}} & K_\mu^{*0} \\ 
K_\mu^{*-} & \bar K_\mu^{*0} & - \sqrt{\frac{2}{3}}\phi_\mu^{(8)}
\end{pmatrix},    
\end{align}
\end{subequations}
where $F_0=(86.2 \pm 0.5)~\rm MeV$ is the pion decay constant in the chiral limit. 
Their chiral transformations are, 
$\Sigma \to \hat L \Sigma \hat R^\dagger$, 
$ B \to \hat h B \hat h^\dagger$, $ V_\mu \to \hat h V_\mu \hat h^\dagger$, 
$\xi \to \hat L \xi \hat h^\dagger = \hat h \xi \hat R^\dagger$, 
where the matrix $\hat h$ is a function of $\hat L,~\hat R$ and $\xi$. 
Note that for $\textsc{X}=B,V_\mu $, under the above chiral transformation:    
$\xi \textsc{X} \xi \to 
\hat L(\xi \textsc{X} \xi) \hat R^\dagger$,  
$\xi^\dagger \textsc{X} \xi^\dagger \to 
\hat R (\xi^\dagger \textsc{X} \xi^\dagger) \hat L^\dagger$, 
$\xi \textsc{X} \xi^\dagger \to 
\hat L (\xi \textsc{X} \xi^\dagger) \hat L^\dagger$,
and $\xi^\dagger \textsc{X} \xi \to 
\hat R (\xi^\dagger \textsc{X} \xi) \hat R^\dagger$. 
For the neutral vector meson $\phi_\mu^{(8)}$, we adopt the ideal mixing in terms of the physical states as $\phi_\mu^{(8)}=\omega_\mu/\sqrt{3}-\phi_\mu \sqrt{2/3}$.

The chiral building blocks consist of the spurion fields, the octet meson and baryon fields and their derivatives. 
To identify the chiral leading terms,
we adopt the chiral power counting scheme in which  
$\{\Sigma,~\xi,~V_\mu,~B,~D_\mu B\}\sim {\cal O}(p^0)$ and 
$\{D_\mu \textsc{X}_\nu,\,D_\mu\Sigma\} \sim {\cal O}(p^1)$.
The covariant derivatives $D_\mu \Sigma$ and $D_\mu \textsc{X}$ ($\textsc{X}=B,\,V_\nu$)
are defined as $D_\mu \Sigma = \partial_\mu \Sigma -il_{\mu}\Sigma+i\Sigma r_{\mu}$, and 
$D_\mu \textsc{X}= \partial_\mu \textsc{X} + [\Gamma_\mu ,\textsc{X}]$, 
where the external sources $l_{\mu}$ and $r_{\mu}$ are traceless matrices in the flavor space and 
$\Gamma_{\mu}=(1/2)\left[\xi(\partial_{\mu}-ir_{\mu})\xi^{\dagger}+\xi^{\dagger}(\partial_{\mu}-il_{\mu})\xi\right]$.
For the irreps $\pmb{6}_{\tL(\tR)}\otimes \pmb{3}_{\tR(\tL)}$ and $\pmb{10}_{\tL(\tR)}\otimes \pmb{1}_{\tR(\tL)}$, 
we need to employ appropriate Lorentz projectors to extract their nontrivial Lorentz irreps correctly: 
$(1,1/2)$ and $(1/2,1)$, and $(3/2,0)$ and $(0,3/2)$. 
We found in Ref.\,\cite{Liao:2025vlj} that, 
for the irreps $\pmb{6}_{\tL(\tR)}\otimes \pmb{3}_{\tR(\tL)}$, the $({1},1/2)$ and $(1/2,{1})$ components of a general vector spinor can be extracted by
$ {\Gamma}_{\mu\nu}^{\tt L,R} \equiv (g_{\mu\nu} - {1\over 4} \gamma_\mu \gamma_\nu)P_{\tt L,R}$, and for the irreps $\pmb{10}_{\tL(\tR)}\otimes \pmb{1}_{\tR(\tL)}$, we apply the following Lorentz projectors~\cite{DelgadoAcosta:2015ypa}:
\begin{align}
\hat\Gamma_{\mu\nu\alpha\beta}^{\tt L,R}  \equiv
{1\over 24} \left( 
2 \{\sigma_{\mu\nu},\sigma_{\alpha\beta}\}
- 
[\sigma_{\mu\nu},\sigma_{\alpha\beta}] 
\right) P_{\tt L,R},
\label{eq:LorentzP}
\end{align}
to extract $(\frac{3}{2},0)$ and $(0,\frac{3}{2})$ components of a tensor spinor. 

Equipped with the above ingredients we studied in Ref.~\cite{Liao:2025vlj} the chiral realizations of the LEFT operators $\calO_i$ by the octet baryons and pseudoscalars in the single-baryon sector, which are collected in SM. It turned out that each operator has one chiral realization at its LO which brings in a low-energy constant (LEC) $c_i$. While two of them have been calculated by lattice QCD~\cite{Yoo:2021gql}, the other two have to be estimated by naive dimensional analysis (NDA)~\cite{Weinberg:1989dx,Manohar:1983md}. Now we do the same thing in the single-baryon and single-vector sector. At the chiral LO of each operator $\calO_i$, the result is 
\begin{align}
{\cal L}_{BV}^{\slashed{B}} &=
d_1 {\rm Tr}\big[ 
{\cal P}_{\bar{\pmb{3}}_\tL \otimes \pmb{3}_\tR}\xi  \gamma_\mu V^\mu B_\tR \xi -
{\cal P}_{\pmb{3}_\tL \otimes \bar{\pmb{3}}_\tR} \xi^\dagger \gamma_\mu V^\mu B_\tL \xi^\dagger 
 \big]
\nonumber
\\
& +d_1' {\rm Tr}\big[ 
{\cal P}_{\bar{\pmb{3}}_\tL \otimes \pmb{3}_\tR}\xi \gamma_\mu B_\tR V^\mu \xi -
{\cal P}_{\pmb{3}_\tL \otimes \bar{\pmb{3}}_\tR}\xi^\dagger \gamma_\mu B_\tL V^\mu \xi^\dagger 
 \big]
\nonumber
\\
& + d_2 {\rm Tr}\big[ 
{\cal P}_{\pmb{8}_\tL \otimes \pmb{1}_\tR}\xi \gamma_\mu V^\mu B_\tR \xi^\dagger
- {\cal P}_{ \pmb{1}_\tL \otimes  \pmb{8}_\tR} \xi^\dagger \gamma_\mu V^\mu B_\tL \xi
\big] 
\nonumber
\\
& + d_2' {\rm Tr}\big[ 
{\cal P}_{\pmb{8}_\tL \otimes \pmb{1}_\tR}\xi \gamma_\mu B_\tR V^\mu  \xi^\dagger
- {\cal P}_{ \pmb{1}_\tL \otimes  \pmb{8}_\tR} \xi^\dagger \gamma_\mu  B_\tL V^\mu\xi
\big] 
\nonumber
\\
& + d_3\big[ {\cal P}_{yzi}^{\tL\tR,\mu}
{\Gamma}_{\mu\nu}^{\tt L} 
(\xi B_\tL \xi)_{yj}
(\xi V^\nu\xi)_{zk} \epsilon_{ijk}
\nonumber
\\
& -{\cal P}_{yzi}^{\tR\tL,\mu}
{\Gamma}_{\mu\nu}^{\tt R}
(\xi^\dagger B_\tR  \xi^\dagger)_{yj}
(\xi^\dagger V^\nu \xi^\dagger)_{kz} \epsilon_{ijk} \big]
\nonumber
\\
& + d_3' \big[ 
{\cal P}_{yzi}^{\tL\tR,\mu}
{\Gamma}_{\mu\nu}^{\tt L} 
(\xi V^\nu B_\tL \xi)_{yj}
\Sigma_{zk} \epsilon_{ijk}
\nonumber
\\
& - {\cal P}_{yzi}^{\tR\tL,\mu}  
{\Gamma}_{\mu\nu}^{\tt R}
(\xi^\dagger V^\nu B_\tR  \xi^\dagger)_{yj}
\Sigma^*_{kz} \epsilon_{ijk} \big]
\nonumber
\\
& +d_3'' \big[ 
{\cal P}_{yzi}^{\tL\tR,\mu} 
{\Gamma}_{\mu\nu}^{\tt L} 
(\xi B_\tL V^\nu \xi)_{yj}
\Sigma_{zk} \epsilon_{ijk}
\nonumber
\\
& - {\cal P}_{yzi}^{\tR\tL,\mu} 
{\Gamma}_{\mu\nu}^{\tt R}
(\xi^\dagger B_\tR V^\nu \xi^\dagger)_{yj}
\Sigma^*_{kz} \epsilon_{ijk} \big]
\nonumber
\\
& + {d_4 \over \Lambda_\chi} \big[ 
{\cal P}_{yzw}^{\tL\tL,\mu\nu} 
\hat\Gamma_{\mu\nu \alpha\beta}^{\tt L} 
(\xi D^{\alpha} B_\tL \xi)_{yi}
\Sigma_{zj}(\xi V^\beta \xi)_{wk} \epsilon_{ijk}
\nonumber
\\
& - {\cal P}_{yzw}^{\tR\tR,\mu\nu} 
\hat\Gamma_{\mu\nu \alpha\beta}^{\tt R}  
(\xi^\dagger D^{\alpha} B_\tR \xi^\dagger)_{yi}
\Sigma^*_{jz} (\xi V^\beta \xi)^*_{kw} 
\epsilon_{ijk}\big]
\nonumber
\\
&+\rm H.c.,
\label{eq:LBlX}
\end{align}%
where $d_{i}$s are unknown LECs.
They are estimated in the SM by NDA to be $d_{1,2,3,4}^{(\prime,\prime\prime)} \sim \Lambda_\chi^2/(4\pi) \approx 0.115\,{\rm GeV}^2$.
Note that except for the irreps $\pmb{10}_{\tL(\tR)}\otimes \pmb{1}_{\tR(\tL)}$ 
all other irreps have more than one chiral realizations at each of their LO. Equation (\ref{eq:LBlX}) is the starting point of all phenomenological analyses on BNV processes in the single-baryon
and single-vector meson sector involving any number of pseudoscalars.

We are now in a position to compare our approach with a formalism developed for the vector meson case over 40 years ago~\cite{Kaymakcalan:1983uc}. 
In that work, the authors treated the vector mesons as the massive Yang-Mills fields of the chiral group 
and extended the results of Ref.\,\cite{Claudson:1981gh} by simply replacing the baryon octet field $B$ by $i\gamma_\mu D^\mu B$ 
to obtain the vector meson Lagrangian for the dim-6 
BNV interactions with $\Delta(B-L)=0$ in $\bar{\pmb{3}}_{\tL(\tR)} \otimes \pmb{3}_{\tR(\tL)}$ and
$\pmb{8}_{\tL(\tR)} \otimes \pmb{1}_{\tR(\tL)}$. The vector mesons enter through the covariant derivative by trading the external sources for them. 
This replacement leads to double counting for the pseudoscalars since it contributes exactly the same pure pseudoscalar BNV Lagrangian terms.
From the EFT perspective, their above operators can be reduced by using the equation of motion for the octet baryon field or its redefinition, to those appearing already in the pseudoscalar case in the absence of the octet vector~\cite{Claudson:1981gh}.
Furthermore, there are additional operators involving covariant derivatives of pseudoscalar mesons that were missed in that work. 
Finally, we incorporated the chiral realizations of the triple-quark operators in the irreps 
$\pmb{6}_{\tL(\tR)} \otimes \pmb{3}_{\tR(\tL)}$ and $\pmb{10}_{\tL(\tR)} \otimes \pmb{1}_{\tR(\tL)}$ first pointed out in~\cite{Liao:2025vlj}.
Our results are fully based on the consistent representation of vector mesons developed in the seminal works~\cite{Weinberg:1968de,Ecker:1989yg}, thereby providing a consistent framework for matching these BNV interactions onto hadronic states involving vector mesons.

\vspace{0.1cm}{\bf Nucleon decays into a vector meson.~}
To apply the formalism developed in \cref{,eq:LBlX} and \cref{eq:LBlM} in the SM to calculate specific decay, it is necessary to identify  relevant quark-level interactions first. As an illustration, we focus on the leading dim-6 BNV operators in the LEFT, as summarized in \cref{tab:dim6ope} in the SM.
In addition to these BNV interactions, the standard ChPT interactions are also required. At LO, the relevant baryon-pseudoscalar (vector) interactions include the conventional $D({\textsc g}_{\textsc d})$ and $F({\textsc g}_{\textsc f})$ terms  \cite{Jenkins:1990jv,Unal:2015hea}, 
\begin{align}
{\cal L}_{\tt ChPT}^{B,0} & \supset
\frac{D}{2} 
{\rm Tr}(\bar B \gamma^\mu \gamma_5\{u_\mu,B\}) 
+ \frac{F}{2} 
{\rm Tr}(\bar B \gamma^\mu \gamma_5 [u_\mu,B])
\notag
\\
& +  {\textsc g}_{\textsc d} 
{\rm Tr}(\bar B \gamma_\mu \{V^\mu, B\}) 
+ {\textsc g}_{\textsc f}  
{\rm Tr}(\bar B \gamma_\mu [V^\mu, B]),
\label{eq:LB0}
\end{align}
where $u_{\mu}=i[\xi(\partial_{\mu}-ir_{\mu})\xi^{\dagger}
-\xi^{\dagger}(\partial_{\mu}-il_{\mu})\xi]$.
For the LECs $D$ and $F$, we adopt the recent lattice 
results $ D=0.730(11)$ and $F=0.447^{6}_{7}$~\cite{Bali:2022qja}. And for the LECs associated with the vector mesons, we use the estimates given in \cite{Unal:2015hea}, ${\textsc g}_{\textsc d}=0$ and ${\textsc g}_{\textsc f}=g/\sqrt{2}$ where the factor $1/\sqrt{2}$ accounts for different parametrization of the interactions. $g$ is the self-coupling of vector mesons and is related to the $\rho$-meson mass and pion decay constant via the well-known Kawarabayashi–Suzuki–Riazuddin–Fayyazuddin relation $g^2=m_\rho^2/(2F_0^2)$~\cite{Kawarabayashi:1966kd,Riazuddin:1966sw}.

\begin{figure}[t]
\centering
\begin{tikzpicture}[mystyle,scale=0.9]
\begin{scope}
\draw[f] (0, 0)node[left]{$\texttt{N}$} -- (2,0);
\draw[f] (2, 0) -- (4,0) node[midway,yshift = + 7pt]{$B$};
\draw[f] (4.0, 0) -- (6,0) node[right]{$l$};
\draw[v, purple] (2,0) -- (2.6,1.3) node[right,yshift = 2 pt]{$X$};
\filldraw [black] (2,0) circle (5pt);
\filldraw [black] (3.86,-0.14) rectangle (4.14,0.14);
\end{scope}
\end{tikzpicture}
\quad\quad
\begin{tikzpicture}[mystyle,scale=0.9]
\begin{scope}
\draw[f] (0, 0)node[left]{$\texttt{N}$} -- (2,0);
\draw[f] (2, 0) -- (4,0) node[right]{$l$};
\draw[v, purple] (2,0) -- (2.6,1.3) node[right,yshift = 2 pt]{$X$};
\filldraw [black] (1.86,-0.14) rectangle (2.14,0.14);
\end{scope}
\end{tikzpicture}
\caption{Diagrams for nucleon BNV two-body decay involving a vector meson in the final state. 
The black square (blob) represents the insertion of a BNV (usual) vertex.}
\label{fig:N2lX_diagram}
\end{figure}
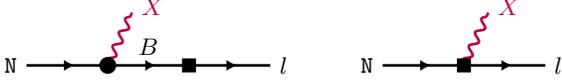

For the nucleon two-body decay modes, ${\tt N}\to lX$, there are two leading-order Feynman diagrams as shown in \cref{fig:N2lX_diagram}.
The relevant interactions are as follows:
\begin{align}
 {\cal L}_{{\tt N} \to l X }   
& = 
C_{{\tt N}\to BX} \bar X^\mu \overline{B}\gamma_\mu{\tt N}
+  C_{Bl}^\tL \overline{l}B_\tL+ C_{Bl}^\tR \overline{l}B_\tR
\nonumber
\\
&+ C_{{\tt N}lX}^\tL \bar X^\mu\overline{l}\gamma_\mu{\tt N}_\tL
+ C_{{\tt N}lX}^\tR \bar X^\mu\overline{l}\gamma_\mu{\tt N}_\tR,
\label{eq:LN2lX}
\end{align}
where the coefficients are summarized in \cref{tab:B2lX_vertex} in the SM for each
process, based on the dim-6 LEFT BNV interactions. 
The decay width is,
\begin{align}
\label{eq:GammaN2lX}
&\Gamma_{{\tt N} \to l X}  = 
\frac{ m_{\tt N} \sqrt{\lambda_X}}{32\pi}
\big[\big(\lambda_X x_X^{-1}
+3(1 + x_l - x_X)\big)
\nonumber
\\
&\times \big( |\tilde C_{{\tt N}lX}^\tL|^2 
+ |\tilde C_{{\tt N}lX}^\tR|^2\big) 
-  12 \sqrt{x_l}
\Re(\tilde C_{{\tt N}lX}^\tL \tilde C_{{\tt N}lX}^{\tR,*}) \big],
\end{align} 
with $\lambda_X=1+x_l^2+x_X^2-2x_l-2x_X-2x_lx_X$,
$x_l = m_l^2/m_{\tt N}^2$, and 
$x_X= m_X^2/m_{\tt N}^2$.
$\tilde C_{{\tt N}lX}^{\tL,\tR}$ are related to the parameters in \cref{eq:LN2lX} via 
\begin{align}
\label{eq:CtNlX}
\tilde C_{{\tt N}lX}^{\tL(\tR)} & 
= C_{{\tt N}lX}^{\tL(\tR)}  
+\sum_B C_{{\tt N}\to BX} 
\frac{m_B C_{Bl}^{\tR(\tL)} + m_l C_{Bl}^{\tL(\tR)}}{m_B^2 - m_l^2},
\end{align}
with a summation over all intermediate baryon $B$. 

\begin{figure}[t]
\centering
\includegraphics[width=0.75\linewidth]{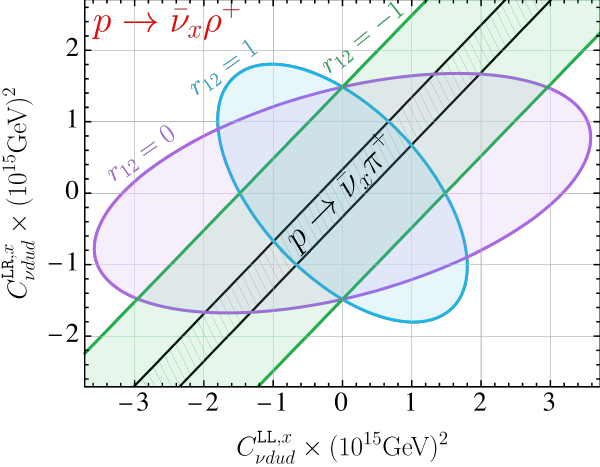}
\vspace{-0.5em}
\caption{An example of vector meson mode in breaking the degeneracy of parameter space in the 
$C_{\nu dud}^{\tL\tL,x}-C_{\nu dud}^{\tL\tR,x}$ plane.}
\label{fig:p2vrho}
\end{figure}

The decay widths for the pseudoscalar modes generally depend on multiple WCs at leading dim-6 order that are modulated by the LECs $c_{1,2}$. Without assuming single-operator dominance, one can only probe an infinite band or cone in the multiple-dimensional parameter space. For convenience and completeness we summarize our results in \cref{app:N2lM} and \cref{tab:B2lM_vertex} in the SM. The decay widths to a vector meson involve the same set of WCs which however are additionally modulated by new LECs $d_{1,2}$. This makes it possible to break the degeneracy with the pseudoscalar modes alone, and constrains the parameter space into a closed region.
For the purpose of illustration, we consider the decays $p\to \bar\nu_x \pi^+$ and $p\to \bar\nu_x \rho^+$, both of which depend on the two WCs, $C_{\nu dud}^{\tL\tR,x}$ and $C_{\nu dud}^{\tL\tL,x}$. 
From \cref{eq:GammaN2lX}, as well as \cref{eq:GammaN2lM} and \cref{tab:B2lX_vertex,tab:B2lM_vertex} in the SM, we have 
\begin{subequations}
\begin{align}
\Gamma_{p \to \bar\nu_x \rho^+} &= 
\frac{m_p (1 - 3 x_\rho^2 + 2 x_\rho^3)}{32\pi\, x_\rho}
 \big( \big| d_1 C_{\nu dud}^{\tL\tR,x} + d_2 C_{\nu dud}^{\tL\tL,x} \big|^2
\nonumber
\\ 
&+{\textsc g}_{\textsc f}^2 m_n^{-2}  
\big|  c_1 C_{\nu dud}^{\tL\tR,x} + c_2 C_{\nu dud}^{\tL\tL,x} \big|^2\big),
\\
\Gamma_{p \to \bar\nu_x \pi^+} & = 
\frac{m_p (1-x_\pi)^2[1 + (D+F)m_p m_n^{-1} ]^2}{64\pi\,F_0^2} 
\nonumber
\\
&\times\big| c_1 C_{\nu dud}^{\tL\tR,x} + c_2 C_{\nu dud}^{\tL\tL,x}\big|^2.
\end{align}
\end{subequations}%
The vector mode indeed depends on the two WCs in a different combination whenever $d_2/d_1\neq c_2/c_1$. 
Once the new LECs $d_{1,2}$ are determined, these two modes can be used jointly to constrain the two WCs. We illustrate this by three benchmark choices for the ratio: $r_{12}\equiv d_1/d_2=1,~0,~-1$ with $d_1=0.115\,\rm GeV^2$. In \cref{fig:p2vrho}, we present the constraints in the $C_{\nu dud}^{\tL\tR,x}-C_{\nu dud}^{\tL\tL,x}$ plane derived from current experimental bounds, assuming both WCs are real. Whereas the $p \to \bar\nu_x \pi^+$ mode restricts the parameter space to a narrow band (hatched gray region), the $p \to \bar\nu_x \rho^+$ mode confines it to an elliptical region 
when $r_{12}\neq-1$. Combining the two constraints reduces the allowed region to a small area, emphasizing the complementary nature of these decay modes. This example highlights the crucial role played by vector meson modes in probing BNV interactions, and underscores the urgent need for a precise determination of these LECs. 

\begin{table}[t]
\centering
\resizebox{\linewidth}{!}{
\renewcommand{\arraystretch}{1.15}
\begin{tabular}{|l|c|l|c|}
\hline 
\multirow{2}*{~~~~Mode}  
& $\Gamma^{-1}_{\rm exp.}$
&\multicolumn{2}{c|}{ Derived new bound }
\\\cline{3-4}
&$~(10^{30}\,\rm yr)~$
& ~~$\Gamma^{-1}_{\rm new} (10^{30}\,\rm yr) $ & LEFT operator
\\
\hline
\multicolumn{4}{|c|}{$\boldsymbol{\Delta(B-L)= 0}$} 
\\
\hline%
$~p\to \bar\nu \rho^+$ & 162   
& $~2.09\times 10^4$
& $\calO_{\nu dud}^{\tL\tR,x},\, \calO_{\nu dud}^{\tL\tL,x}$
\\
\hline%
&  
& $~6.04\times 10^7$ 
& $\calO_{\nu dsu}^{\tL\tR,x},\, \calO_{\nu dsu}^{\tL\tL,x}$
\\
$~p\to \bar\nu K^{*+}$ & 51 
& $~9.67\times 10^5$ 
& $\calO_{\nu sud}^{\tL\tR,x},\, \calO_{\nu sud}^{\tL\tL,x}$
\\
& 
& $~2.85\times 10^5$ 
& $\calO_{\nu uds}^{\tL\tR,x}\,\qquad\quad$
\\
\hline%
$~p\to e^+ \rho^0 $ & 720 
& $~9.10\times 10^5$ 
& $\calO_{\ell uud}^{\tL\tR/\tR\tL,e},\, \calO_{\ell uud}^{\tL\tL/\tR\tR,e}$
\\
\hline%
$~p\to \mu^+ \rho^0 $ & 570 
& $~1.37\times 10^6$ 
& $\calO_{\ell uud}^{\tL\tR/\tR\tL,\mu},\,\calO_{\ell uud}^{\tL\tL/\tR\tR,\mu}$
\\
\hline%
$~p\to e^+ \omega $ & 1600 
& $~9.98\times 10^5$ 
& $\calO_{\ell uud}^{\tL\tR/\tR\tL,e},\,\calO_{\ell uud}^{\tL\tL/\tR\tR,e}$
\\
\hline%
$~p\to \mu^+ \omega $ & 2800 
& $~1.57\times 10^6$ 
& $\calO_{\ell uud}^{\tL\tR/\tR\tL,\mu},\,\calO_{\ell uud}^{\tL\tL/\tR\tR,\mu}$
\\
\hline%
$~p\to e^+ K^{*0}~$ & 84 
& $~1.50(0.59)\times 10^5~$ 
& $~\calO_{\ell usu}^{\tL\tR/\tR\tL,e}\,(\calO_{\ell usu}^{\tL\tL/\tR\tR,e})~$
\\
\hline%
$~p\to e^+ \bar K^{*0} $ & 0.6 
& \multicolumn{2}{c|}{\ding{55}\,(weak/5-quark operators)} 
\\
\hline
$~n\to e^+ \rho^- $ &  217  
& $~4.48\times 10^5$ 
& $\calO_{\ell uud}^{\tL\tR/\tR\tL,e},\, \calO_{\ell uud}^{\tL\tL/\tR\tR,e}$
\\
\hline%
$~n\to \mu^+ \rho^- $ & 228 
& $~6.69\times 10^5$ 
& $\calO_{\ell uud}^{\tL\tR/\tR\tL,\mu},\, \calO_{\ell uud}^{\tL\tL/\tR\tR,\mu}$
\\
\hline%
$~n\to e^+ K^{*-} $~ & 0.6   
& \multicolumn{2}{c|}{\ding{55}\,(weak/5-quark operators)} 
\\
\hline%
$~n\to \bar\nu \rho^0 $ & 19  
& $~4.13\times 10^4$ 
& $\calO_{\nu dud}^{\tL\tR,x},\, \calO_{\nu dud}^{\tL\tL,x}$
\\
\hline%
$~n\to \bar\nu \omega $ & 108 
& $~4.53\times 10^4$ 
& $\calO_{\nu dud}^{\tL\tR,x},\, \calO_{\nu dud}^{\tL\tL,x}$
\\
\hline%
& 
& $~8.46\times 10^4$ 
& $\calO_{\nu dsu}^{\tL\tR,x},\, \calO_{\nu dsu}^{\tL\tL,x}$
\\
$~n\to \bar\nu K^{*0}$ & 78 
& $~1.22\times 10^6$ 
& $\calO_{\nu sud}^{\tL\tR,x},\, \calO_{\nu sud}^{\tL\tL,x}$
\\
& 
& $~3.23\times 10^8$ 
& $\calO_{\nu uds}^{\tL\tR,x}\,\qquad\quad$
\\
\hline
\multicolumn{4}{|c|}{\cellcolor{gray!15}$\boldsymbol{\Delta(B+L)=0}$}
\\
\hline%
$~p\to \nu \rho^+$ & 162 
& $~2.09\times 10^4$
& $\calO_{\bar\nu dud}^{\tR\tL,x},\, \calO_{\bar\nu dud}^{\tR\tR,x}$
\\
\hline%
&  
& $~6.04\times 10^7$ 
& $\calO_{\bar\nu dsu}^{\tR\tL,x},\, \calO_{\bar\nu dsu}^{\tR\tR,x}$
\\
$~p\to \nu K^{*+}$ & 51
& $~9.67\times 10^5$ 
& $\calO_{\bar\nu sud}^{\tR\tL,x},\, \calO_{\bar\nu sud}^{\tR\tR,x}$
\\
& 
& $~2.84\times 10^5$ 
& $\calO_{\bar\nu uds}^{\tR\tL,x}\,\qquad\quad$
\\\hline
$~n\to e^- \rho^+ $ &  62
& \multicolumn{2}{c|}{\multirow{2}{*}{\ding{55}\,($\Delta I=3/2$)}}
\\
$~n\to \mu^- \rho^+ $ & 7 & \multicolumn{2}{c|}{}
\\
\hline
$~n\to e^- K^{*+} $ & $-$ & $~3.10 (1.61) \times 10^3$ & $~\calO_{\ell dds}^{\tL\tR/\tR\tL,e}\,(\calO_{\ell dds}^{\tL\tL/\tR\tR,e})~$
\\
\hline%
$~n\to \nu \rho^0 $ & 19 
& $~4.13\times 10^4$ 
& $\calO_{\bar\nu dud}^{\tR\tL,x},\, \calO_{\bar\nu dud}^{\tR\tR,x}$
\\
\hline%
$~n\to \nu \omega $ & 108
& $~4.53\times 10^4$ 
& $\calO_{\bar\nu dud}^{\tR\tL,x},\, \calO_{\bar\nu dud}^{\tR\tR,x}$
\\
\hline%
& 
& $~8.46\times 10^4$ 
& $\calO_{\bar\nu dsu}^{\tR\tL,x},\, \calO_{\bar\nu dsu}^{\tR\tR,x}$
\\
$~n\to \nu K^{*0}$ & 78 
& $~1.22\times 10^6$ 
& $\calO_{\bar\nu sud}^{\tR\tL,x},\, \calO_{\bar\nu sud}^{\tR\tR,x}$
\\
& 
& $~3.23\times 10^8$ 
& $\calO_{\bar\nu uds}^{\tR\tL,x}\,\qquad\quad$
\\\hline
\end{tabular} }
\caption{Derived new bounds on the nucleon two-body decay modes involving a vector meson based on those of the corresponding two-body pseudoscalar modes, under the assumption of single operator dominance.
The modes marked with a \ding{55} are suppressed by additional weak interactions or dim-9 interactions and are therefore neglected. 
}
\label{tab:N2lXbound}
\end{table}

As a second exercise, we neglect the unknown LECs $d_i$s and derive bounds on the vector meson modes by leveraging the experimental limits on the pseudoscalar modes. Assuming single operator dominance, the branching ratio of a vector meson mode is directly related to that of its corresponding pseudoscalar mode. In \cref{tab:N2lXbound}, we present the derived bounds along with the associated bridging operators. For comparison, the second column lists the direct experimental lower bounds on their inverse decay widths 
\cite{ParticleDataGroup:2024cfk}. 
As can be seen, for all kinematically allowed two-body vector meson modes (except $p\to e^+ \bar K^{*0}$ and $n\to e^+ K^{*-},~e^-\rho^+,~\mu^-\rho^+$), the obtained indirect bounds are stronger than the direct experimental limits by 
2--6 orders of magnitude. 
In particular, for the processes $p\to \nu(\bar\nu) K^{*+}$ and $n\to \nu(\bar\nu) K^{*0}$, the much stronger bounds associated with the respective operators $\calO^{\tL\tR/\tL\tL,x}_{\nu dsu}(\calO^{\tR\tL/\tR\tR,x}_{\nu dsu})$ and $\calO^{\tL\tR,x}_{\nu uds}(\calO^{\tR\tL,x}_{\bar\nu uds})$ are due to cancellations between the $\Sigma^0$- and $\Lambda^0$-mediated contributions. 
Last but not least, for the experimentally unconstrained mode $n\to e^- K^{*+}$, 
we have derived a stringent upper bound based on that of the pseudoscalar mode $n\to e^- K^+$.

\vspace{0.1cm}{\bf Summary.~}
In this work, we developed a consistent chiral framework for nucleon decays involving vector mesons. Starting from the general triple-quark operator structures in the LEFT \cite{Jenkins:2017jig}, 
we performed their LO chiral realizations, treating both vector-meson and baryon octets as matter fields with well-defined chiral transformation properties. The relevant low-energy constants were estimated using NDA. To highlight the phenomenological significance of vector meson modes, we demonstrated through examples that including these modes helps resolve degeneracies in the parameter space that cannot be constrained solely by pseudoscalar modes. Moreover, by employing the stringent constraints on the dim-6 WCs derived from the pseudoscalar modes, we established tighter constraints on vector meson modes than those currently available from experiments, when considering only the long-distance contributions. 

Should a BNV signal be observed, the presence of these additional vector meson decay channels would provide valuable information for identifying the underlying interaction structures responsible for the signal. For specific new physics scenarios, our formalism serves as a bridge connecting different decay modes and predicts correlations among them that can be tested in experimental searches. In addition, our formalism is not limited to the usual nucleon decays, but also covers scenarios involving light particles in the final state when the spurion fields are accordingly chosen, as well as BNV decays of the tau lepton and heavier hyperons. A comprehensive study of all relevant decay channels thus enables us to probe a broader parameter space and gain deeper insights into fundamental questions such as baryogenesis, neutrino masses, and the nature of dark matter.

\section*{Acknowledgements}
This work was supported 
by the Grants 
No.\,NSFC-12305110,
No.\,NSFC-12247151, 
and No.\,NSFC-12035008. 

\bibliography{references.bib}{}
\bibliographystyle{utphys}

\onecolumngrid
\newpage
\setcounter{table}{0}
\renewcommand{\thetable}{S\arabic{table}}
\setcounter{equation}{0}
\renewcommand{\theequation}{S\arabic{equation}}
\begin{center}
\vspace{0.3cm}{\Large\bf Supplemental Material}
\end{center}

\section{Chiral Lagrangian for BNV nucleon decay with octet pseudoscalar mesons}
\label{app:BNVChPTpseudoscalar}

For simplicity, we organize the spurion fields associated with the irreps ${\cal N}_{\pmb{8}_\tL \otimes \pmb{1}_\tR}$ and ${\cal N}_{\bar{\pmb{3}}_\tL \otimes \pmb{3}_\tR}$ in matrix form, 
\begin{align}
{\cal P}_{\pmb{8}_\tL \otimes \pmb{1}_\tR}  = 
 \begin{pmatrix}
0   & {\cal P}_{dds}^{\tL\tL}  
& {\cal P}_{sds}^{ \tL\tL} 
\\[1pt]
{\cal P}_{usu}^{\tL\tL}  
& {\cal P}_{dsu}^{\tL\tL} 
& {\cal P}_{ssu}^{\tL\tL} 
\\[1pt] 
{\cal P}_{uud}^{\tL\tL}  
& {\cal P}_{dud}^{\tL\tL} 
& {\cal P}_{sud}^{\tL\tL}
 \end{pmatrix}, 
\quad 
{\cal P}_{\pmb{3}_\tL \otimes \bar{\pmb{3}}_\tR }  = 
 \begin{pmatrix}
{\cal P}_{uds}^{\tR\tL}  
&  {\cal P}_{dds}^{\tR\tL} 
& {\cal P}_{sds}^{\tR\tL} 
\\[1pt]
{\cal P}_{usu}^{\tR\tL}  
&  {\cal P}_{dsu}^{\tR\tL} 
& {\cal P}_{ssu}^{\tR\tL} 
\\[1pt] 
{\cal P}_{uud}^{\tR\tL}  
&  {\cal P}_{dud}^{\tR\tL} 
& {\cal P}_{sud}^{\tR\tL}
\end{pmatrix}. 
\end{align}
Note that we have applied the condition $\textrm{Tr}{\cal N}_{{\bf 8}_\tL\otimes {\bf 1}_\tR}=0$ to attribute the $(1,1)$ entry of ${\cal P}_{{\bf 8}_\tL\otimes {\bf 1}_\tR}$ to its $(2,2)$ and $(3,3)$ entries by treating ${\cal N}^{\tL\tL}_{uds}$ as redundant~\cite{Liao:2025vlj}.
For the triple-quark interactions in \cref{eq:q3LEFT}, the LO chiral realizations involving octet pseudoscalar mesons and baryons are~\cite{Liao:2025vlj},
\begin{align}
{\cal L}_{B}^{\slashed{B}} &=
c_1 {\rm Tr}\big[ 
{\cal P}_{  \bar{\pmb{3}}_\tL \otimes \pmb{3}_\tR} \xi B_\tL \xi -
{\cal P}_{\pmb{3}_\tL \otimes \bar{\pmb{3}}_\tR} \xi^\dagger B_\tR \xi^\dagger 
 \big]
 + c_2 {\rm Tr}\big[ 
{\cal P}_{\pmb{8}_\tL \otimes \pmb{1}_\tR}\xi B_\tL \xi^\dagger
- {\cal P}_{ \pmb{1}_\tL \otimes  \pmb{8}_\tR} \xi^\dagger B_\tR \xi
\big] 
\nonumber
\\
& + {c_3 \over \Lambda_\chi} \big[ 
{\cal P}_{yzi}^{\tL\tR,\mu}
{\Gamma}_{\mu\nu}^{\tt L} 
(\xi i D^\nu B_\tL \xi)_{yj}
\Sigma_{zk} \epsilon_{ijk}
- {\cal P}_{yzi}^{\tR\tL,\mu}
{\Gamma}_{\mu\nu}^{\tt R}
(\xi^\dagger i D^\nu B_\tR  \xi^\dagger)_{yj}
\Sigma^*_{kz} \epsilon_{ijk} \big]
\nonumber
\\
&
{c_4 \over \Lambda_\chi^2} \big[ 
{\cal P}_{yzw}^{\tL\tL,\mu\nu}
\hat\Gamma_{\mu\nu \alpha\beta}^{\tt L} 
(\xi D^{\alpha} B_\tL \xi)_{yi}
\Sigma_{zj}(D^{\beta}\Sigma)_{wk} \epsilon_{ijk}
- {\cal P}_{yzw}^{\tR\tR,\mu\nu} 
\hat\Gamma_{\mu\nu \alpha\beta}^{\tt R}  
(\xi^\dagger D^{\alpha} B_\tR \xi^\dagger)_{yi}
\Sigma^*_{jz} (D^{\beta}\Sigma)^*_{kw} \epsilon_{ijk}\big]
+\hc.
\label{eq:LBlM}
\end{align}%
where the $c_{1,2,3,4}$ are the LECs. The LECs $c_{1,2}$ can be identified with $\alpha$ and $\beta$ used in the literature, and the recent LQCD results lead to $c_1=\alpha=-0.01257(111)\,{\rm GeV}^3$ and $c_2=\beta=0.01269(107)\,{\rm GeV}^3$ \cite{Yoo:2021gql}, respectively.
The other two LECs $c_{3,4}$ have been estimated using the NDA method~\cite{Weinberg:1989dx,Manohar:1983md}, which yields $c_3 \sim \Lambda_\chi^3/(4\pi)^2 \approx 0.011\,{\rm GeV}^3$
and 
$c_4 \sim \Lambda_\chi^2 F_0 /[\sqrt{2} (4\pi)] \approx 0.007\,{\rm GeV}^3$ \cite{Liao:2025vlj}.

\section{Naive dimensional analysis estimation of the new LECs}

The same NDA method mentioned above can be used to determine the new LECs associated with the BNV hadronic interactions involving vector mesons.
Following Weinberg's approach~\cite{Weinberg:1989dx}, we introduce reduced couplings for each quark-level operator and its hadron-level realizations, and then match them to determine the LEC for each realization. For an interaction term involving a coupling constant $\tilde g$ and a dim-$D$ operator that contains a minimum number $m$ of physical fields, its reduced coupling is defined as $\tilde g(4\pi)^{2-m} \Lambda_\chi^{D-4}$. 
Since the spurion fields act as a spectator in chiral matching, it is sufficient to consider the pure triple-quark factor in \cref{eq:LEFT3qO}, for which we have $\tilde g=1$, $m=3$, and $D=9/2$, resulting in a reduced coupling given by $C_q = (4\pi)^{-1} \Lambda_\chi^{1/2}$. For the corresponding hadron-level realizations in \cref{eq:LBlX}, we expand the meson matrix to the first nonvanishing order and obtain $d_{1,2,3}^{(\prime,\prime\prime)} B X$ ($\tilde g=d_{1,2,3}^{(\prime,\prime\prime)}$, $m=2$, and $D=5/2$), and  $(d_4/\Lambda_\chi) \partial B X$ ($\tilde g=d_4/\Lambda_\chi$, $m=2$, and $D=7/2$).
Consequently, we find the reduced couplings for the hadron-level realizations to be 
$C_{1,2,3,4} =d_{1,2,3,4}^{(\prime,\prime\prime)} \Lambda_\chi^{-3/2} $.
By identifying $C_{1-4}=C_q$, 
we ultimately obtain  
$d_{1,2,3,4}^{(\prime,\prime\prime)} \sim \Lambda_\chi^2/(4\pi) \approx 0.115\,{\rm GeV}^2$.

\section{Effective BNV interaction vertices arising from dim-6 LEFT interactions}
\label{app:hadronicLag}

\begin{table}[h]
	\center
    \resizebox{0.6\linewidth}{!}{
    \renewcommand{\arraystretch}{1.5}
	\begin{tabular}{|c|c|c|c|}
		\hline
        \multicolumn{2}{|c|}{$\boldsymbol{\Delta(B-L)= 0}$} &
		\multicolumn{2}{|c|}{\cellcolor{gray!15}$\boldsymbol{\Delta(B+L)=0}$}
		\\
		\hline
		~~~$\calO_{\nu dud}^{\tL\tL}$~~~  & 
		~~~$(\overline{\nu_{\tL}^{\C}} d_{\tL}^\alpha) 
        (\overline{u_{\tL}^{\beta \C}} d_{\tL}^\gamma)
        \epsilon_{\alpha \beta \gamma}$~~~  &
		~~~$\calO_{\bar{\ell} ddd}^{\tL\tL}$~~~  &
		~~~$(\overline{\ell_{\tR}} d_{\tL}^\alpha) 
        (\overline{d_{\tL}^{\beta \C}} d_{\tL}^\gamma)
        \epsilon_{\alpha \beta \gamma}$ ~~~
		\\
		\hline
		$\calO_{\ell udu}^{\tL\tL}$  & 
		$(\overline{\ell_{\tL}^{\C}} u_{\tL}^\alpha) 
        (\overline{d_{\tL}^{\beta \C}} u_{\tL}^\gamma)
        \epsilon_{\alpha \beta \gamma}$  &
		$\calO_{\bar{\nu} dud}^{\tR\tL}$  &
		$(\overline{\nu_{\tL}} d_{\tR}^\alpha) 
        (\overline{u_{\tL}^{\beta \C}} d_{\tL}^\gamma)
        \epsilon_{\alpha \beta \gamma}$ 
		\\
		\hline
		$\calO_{\ell duu}^{\tR\tL}$  & 
		$(\overline{\ell_{\tR}^{\C}} d_{\tR}^\alpha) 
        (\overline{u_{\tL}^{\beta \C}} u_{\tL}^\gamma)
        \epsilon_{\alpha \beta \gamma}$  &
		$\calO_{\bar{\nu} udd}^{\tR\tL}$  &
		$(\overline{\nu_{\tL}} u_{\tR}^\alpha) 
        (\overline{d_{\tL}^{\beta \C}} d_{\tL}^\gamma)
        \epsilon_{\alpha \beta \gamma}$ 
		\\
		\hline
		$\calO_{\ell udu}^{\tR\tL}$  & 
		$(\overline{\ell_{\tR}^{\C}} u_{\tR}^\alpha) 
        (\overline{d_{\tL}^{\beta \C}} u_{\tL}^\gamma)
        \epsilon_{\alpha \beta \gamma}$  &
		$\calO_{\bar{\ell} ddd}^{\tR\tL}$  &
		$(\overline{\ell_{\tL}} d_{\tR}^\alpha) 
        (\overline{d_{\tL}^{\beta \C}} d_{\tL}^\gamma)
        \epsilon_{\alpha \beta \gamma}$ 
		\\
		\hline
		$\calO_{\ell duu}^{\tL\tR}$  & 
		$(\overline{\ell_{\tL}^{\C}} d_{\tL}^\alpha) 
        (\overline{u_{\tR}^{\beta \C}} u_{\tR}^\gamma)
        \epsilon_{\alpha \beta \gamma}$  &
		$\calO_{\bar{\ell} ddd}^{\tL\tR}$  &
		$(\overline{\ell_{\tR}} d_{\tL}^\alpha) 
        (\overline{d_{\tR}^{\beta \C}} d_{\tR}^\gamma)
        \epsilon_{\alpha \beta \gamma}$ 
		\\
		\hline
		$\calO_{\ell udu}^{\tL\tR}$  &  
		$(\overline{\ell_{\tL}^{\C}} u_{\tL}^\alpha) 
        (\overline{d_{\tR}^{\beta \C}} u_{\tR}^\gamma)
        \epsilon_{\alpha \beta \gamma}$  &
		$\calO_{\bar{\nu} dud}^{\tR\tR}$  &
		$(\overline{\nu_{\tL}} d_{\tR}^\alpha) 
        (\overline{u_{\tR}^{\beta \C}} d_{\tR}^\gamma)
        \epsilon_{\alpha \beta \gamma}$ 
		\\
		\hline
		$\calO_{\nu ddu}^{\tL\tR}$  & 
		$(\overline{\nu_{\tL}^{\C}} d_{\tL}^\alpha) 
        (\overline{d_{\tR}^{\beta \C}} u_{\tR}^\gamma)
        \epsilon_{\alpha \beta \gamma}$  &
		$\calO_{\bar{\ell} ddd}^{\tR\tR}$  &
		$(\overline{\ell_{\tL}} d_{\tR}^\alpha) 
        (\overline{d_{\tR}^{\beta \C}} d_{\tR}^\gamma)
        \epsilon_{\alpha \beta \gamma}$ 
		\\
		\hline
		$\calO_{\nu udd}^{\tL\tR}$  & 
		$(\overline{\nu_{\tL}^{\C}} u_{\tL}^\alpha) 
        (\overline{d_{\tR}^{\beta \C}} d_{\tR}^\gamma)
        \epsilon_{\alpha \beta \gamma}$  &  &
		\\
		\cline{1-2}
		$\calO_{\ell udu}^{\tR\tR}$  & 
		$(\overline{\ell_{\tR}^{\C}} u_{\tR}^\alpha) 
        (\overline{d_{\tR}^{\beta \C}} u_{\tR}^\gamma)
        \epsilon_{\alpha \beta \gamma}$  &  &
		\\
		\hline
	    \end{tabular}
    }
\caption{The LEFT dim-6 operators with $\Delta B=1$ and $\Delta L=\pm 1$. $\alpha,\beta,\gamma$ are color indices while the flavor indices are omitted for simplicity. }
\label{tab:dim6ope}
\end{table}

The leading dim-6 operators in the LEFT with either $\Delta(B-L)=0$ or $\Delta (B+L)=0$ are summarized in \cref{tab:dim6ope}, adapted from 
\cite{Jenkins:2017jig} with minor notation modifications.
The lepton components multiplied by their WCs lead to the following spurion matrices \cite{Fan:2024gzc}:
\begin{subequations}
\label{eq:spu}
\begin{align}
{\cal P}_{\pmb{8}_\tL \otimes \pmb{1}_\tR} = 
\begin{pmatrix}
0  & {\cellcolor{gray!15}{C^{\tL\tL,x}_{\bar{\ell} dds} \overline{\ell_{\tR x}}}}
&  {\cellcolor{gray!15}C^{\tL\tL,x}_{\bar{\ell} sds} \overline{\ell_{\tR x}} }
\\[3pt]
C^{\tL\tL,x}_{\ell usu} \overline{\ell_{\tL x}^{\C}} 
& C^{\tL\tL,x}_{\nu dsu} \overline{\nu_{\tL x}^{\C}} 
& C^{\tL\tL,x}_{\nu ssu} \overline{\nu_{\tL x}^{\C}} 
\\[3pt]
C^{\tL\tL,x}_{\ell uud} \overline{\ell_{\tL x}^{\C}} 
& C^{\tL\tL,x}_{\nu dud} \overline{\nu_{\tL x}^{\C}} 
& C^{\tL\tL,x}_{\nu sud} \overline{\nu_{\tL x}^{\C}} 
\end{pmatrix},&
\quad
{\cal P}_{\pmb{1}_\tL \otimes \pmb{8}_\tR} = 
\begin{pmatrix}
0  & {\cellcolor{gray!15}C^{\tR\tR,x}_{\bar{\ell} dds} \overline{\ell_{\tL x}}}
&  {\cellcolor{gray!15}C^{\tR\tR,x}_{\bar{\ell} sds} \overline{\ell_{\tL x}}} \\[3pt]
C^{\tR\tR,x}_{\ell usu} \overline{\ell_{\tR x}^{\C}} 
&  {\cellcolor{gray!15}C^{\tR\tR,x}_{\bar{\nu} dsu}  \overline{\nu_{\tL x}}} 
&  {\cellcolor{gray!15}C^{\tR\tR,x}_{\bar{\nu} ssu}  \overline{\nu_{\tL x}}}  \\[3pt]
C^{\tR\tR,x}_{\ell uud} \overline{\ell_{\tR x}^{\C}} 
&  {\cellcolor{gray!15}C^{\tR\tR,x}_{\bar{\nu} dud}  \overline{\nu_{\tL x}} }
&  {\cellcolor{gray!15}C^{\tR\tR,x}_{\bar{\nu} sud} \overline{\nu_{\tL x}} }
\end{pmatrix},
\\
{\cal P}_{\pmb{3}_\tL \otimes \bar{\pmb{3}}_\tR} = 
\begin{pmatrix}	
{\cellcolor{gray!15}C^{\tR\tL,x}_{\bar{\nu}uds}\overline{\nu_{\tL x}}  } 
& {\cellcolor{gray!15}C^{\tR\tL,x}_{\bar{\ell} dds} \overline{\ell_{\tL x}} }
& {\cellcolor{gray!15}C^{\tR\tL,x}_{\bar{\ell} sds} \overline{\ell_{\tL x}}} \\[3pt]
C^{\tR\tL,x}_{\ell usu} \overline{\ell_{\tR x}^{\C}}
& {\cellcolor{gray!15}C^{\tR\tL,x}_{\bar{\nu} dsu}  \overline{\nu_{\tL x}} }
&  {\cellcolor{gray!15}C^{\tR\tL,x}_{\bar{\nu} ssu}  \overline{\nu_{\tL x}}} \\[3pt]
C^{\tR\tL,x}_{\ell uud} \overline{\ell_{\tR x}^{\C}} 
&{\cellcolor{gray!15}C^{\tR\tL,x}_{\bar{\nu} dud}  \overline{\nu_{\tL x}} }
&{\cellcolor{gray!15}C^{\tR\tL,x}_{\bar{\nu} sud} \overline{\nu_{\tL x}} }
\end{pmatrix},&
\quad
{\cal P}_{\bar{\pmb{3}}_\tL \otimes \pmb{3}_\tR} = 
\begin{pmatrix}
C^{\tL\tR,x}_{\nu uds}\overline{\nu_{\tL x}^{\C}} &
{\cellcolor{gray!15}C^{\tL\tR,x}_{\bar{\ell} dds} \overline{\ell_{\tR x}} }
&{\cellcolor{gray!15}{C^{\tL\tR,x}_{\bar{\ell}sds} \overline{\ell_{\tR x}}} } \\[3pt]
C^{\tL\tR,x}_{\ell usu} \overline{\ell_{\tL x}^{\C}}
& C^{\tL\tR,x}_{\nu dsu}  \overline{\nu_{\tL x}^{\C}} 
&C^{\tL\tR,x}_{\nu ssu}  \overline{\nu_{\tL x}^{\C}} 
\\[3pt]
C^{\tL\tR,x}_{\ell uud} \overline{\ell_{\tL x}^{\C}} 
& C^{\tL\tR,x}_{\nu dud}  \overline{\nu_{\tL x}^{\C}} 
&C^{\tL\tR,x}_{\nu sud} \overline{\nu_{\tL x}^{\C}} 
\end{pmatrix}.
\end{align}
\end{subequations}
We do not strictly follow the notation used in \cref{tab:dim6ope}. Instead, we explicitly denote the quark flavors as subscripts in the WCs, and slightly reorder the positions of the two quark fields in the second fermion bilinear of each operator and its corresponding WC to maintain consistency with the chiral representations. 

By substituting the spurion matrices from \cref{eq:spu} into \cref{eq:LBlM} and expanding the pseudoscalar matrix to the zeroth order in the meson fields, we obtain the following baryon-lepton mass mixing terms ($\mathcal{L}_{ Bl}$),
\begin{subequations}
\begin{align}
{\cal L}_{Bl}^{{\tt \Delta( B-L)=0}} & = 
	\left(c_1 C^{\tL \tR,x}_{\ell uud} + 
  c_2C^{\tL \tL,x}_{\ell uud } \right)
    ( \overline{\ell_{\tL x}^{\C}} p_{\tL} ) 
    -\left( c_1 C^{\tR \tL,x}_{\ell uud} + 
  c_2C^{\tR \tR,x}_{\ell uud} \right)
    ( \overline{\ell_{\tR x}^{\C}} p_{\tR} )
    \notag\\
	&\quad + \left(c_1 C^{\tL \tR,x}_{\ell usu} + 
  c_2C^{\tL \tL,x}_{\ell usu} \right)
    ( \overline{\ell_{\tL x}^{\C}} \Sigma^+_{\tL} )
    -\left(c_1 C^{\tR \tL,x}_{\ell usu} + 
  c_2C^{\tR \tR,x}_{\ell usu} \right)
    ( \overline{\ell_{\tR x}^\C} \Sigma^+_{\tR} ) 
    \notag\\
    &\quad +\frac{1}{\sqrt{2}} \left[c_1 (C^{\tL \tR,x}_{\nu uds} - C^{\tL \tR,x}_{\nu dsu})
    - c_2 C^{\tL \tL,x}_{\nu dsu} \right]
    ( \overline{\nu_{\tL x}^{\C}} \Sigma^0_{\tL} )  
    \notag\\
	&\quad +\frac{1}{\sqrt{6}}\left[c_1 
    (C^{\tL \tR,x}_{\nu uds} + C^{\tL \tR,x}_{\nu dsu} 
    - 2 C^{\tL \tR,x}_{\nu sud} )
    + c_2 ( C^{\tL \tL,x}_{\nu dsu}  - 2 C^{\tL \tL,x}_{\nu sud}) \right] 
    ( \overline{\nu_{\tL x}^{\C}} \Lambda^0_{\tL} )
	  \notag\\
	&\quad +\left(c_1 C^{\tL \tR,x}_{\nu dud} + 
  c_2C^{\tL \tL,x}_{\nu dud} \right)
    ( \overline{\nu_{\tL x}^{\C}} n_{\tL} )
    +\left(c_1 C^{\tL \tR,x}_{\nu ssu} + 
  c_2C^{\tL \tL,x}_{\nu ssu} \right)
    ( \overline{\nu^{\C}_{\tL x}} \Xi^0_{\tL} )
    +\hc, 
\\
{\cal L}_{Bl}^{{\tt \Delta( B+L)=0}} & = 
   \left(c_1 C^{\tL \tR,x}_{\bar{\ell}sds} + 
  c_2 C^{\tL \tL,x}_{\bar{\ell}sds} \right)
    ( \overline{\ell_{\tR x}} \Xi^-_{\tL} ) 
	-\left(c_1 C^{\tR \tL,x}_{\bar{\ell}sds} + 
  c_2C^{\tR \tR,x}_{\bar{\ell}sds } \right)
    ( \overline{\ell_{\tL x}} \Xi^-_{\tR} )  
    \notag\\
    &\quad + \left(c_1 C^{\tL \tR,x}_{\bar{\ell} dds} + 
  c_2 C^{\tL \tL,x}_{\bar{\ell} dds} \right)
    ( \overline{\ell_{\tR x}} \Sigma^-_{\tL} ) 
    -\left(c_1 C^{\tR \tL,x}_{\bar{\ell} dds} + 
  c_2 C^{\tR \tR,x}_{\bar{\ell} dds} \right) 
   ( \overline{\ell_{\tL x}} \Sigma^-_{\tR} ) 
    \notag\\
	&\quad - \frac{1}{\sqrt{2}} \left[c_1 (C^{\tR \tL,x}_{\bar{\nu}uds} - C^{\tR \tL,x}_{\bar{\nu}dsu}) - 
  c_2 C^{\tR \tR,x}_{\bar{\nu}dsu} \right]
    ( \overline{\nu_{\tL x}} \Sigma^0_{\tR} ) 
    \notag\\
	&\quad - \frac{1}{\sqrt{6}}\left[c_1 (C^{\tR \tL,x}_{\bar{\nu}uds} 
    +  C^{\tR \tL,x}_{\bar{\nu}dsu} - 2 C^{\tR \tL,x}_{\bar{\nu} sud} )+ 
  c_2 (C^{\tR \tR,x}_{\bar{\nu}dsu}  - 2 C^{\tR \tR,x}_{\bar{\nu} sud} ) \right]
    ( \overline{\nu_{\tL x}} \Lambda^0_{\tR} )
	\notag\\
    &\quad -\left(c_1 C^{\tR \tL,x}_{\bar{\nu}dud} + 
  c_2 C^{\tR \tR,x}_{\bar{\nu}dud} \right)
    ( \overline{\nu_{\tL x}} n_{\tR} ) 
    -\left(c_1 C^{\tR \tL,x}_{\bar{\nu}ssu} + 
  c_2 C^{\tR \tR,x}_{\bar{\nu}ssu} \right)
    ( \overline{\nu_{\tL x}} \Xi^0_{\tR} )   
    +\hc, 
\end{align} 
\end{subequations}
where the superscript or subscript $x$ is a lepton flavor index. 
Similarly, by inserting the explicit form of the spurion fields into \cref{eq:LBlX} and expanding the pseudoscalar matrix to the zeroth order in the meson fields, we obtain the following three-point interactions involving a nucleon and a vector meson ($\mathcal{L}_{\texttt{N}lX}$),
\begin{subequations}
\begin{align}
\nonumber
{\cal L}_{{\tt N}lX}^{{\tt \Delta( B-L)=0}} & =  
\frac{1}{\sqrt{2}} \left\{(d_1 C^{\tL\tR,x}_{\ell uud} + d_2 C^{\tL\tL,x}_{\ell uud}  ) (\overline{\ell_{\tL x}^\C} \gamma^\mu p_\tR) - (d_1 C^{\tR\tL,x}_{\ell uud} + d_2 C^{\tR\tR,x}_{\ell uud}  ) (\overline{\ell_{\tR x}^\C} \gamma^\mu p_\tL)\right\} \rho^0_\mu 
\\
\nonumber
&\quad + \frac{1}{\sqrt{6}} \left\{\left[(d_1 - 2 d_1^\prime) C_{\ell uud}^{\tL\tR,x} + 
(d_2- 2 d_2^\prime) C_{\ell uud}^{\tL\tL,x} \right] (\overline{\ell_{\tL x}^\C} \gamma^\mu p_\tR)
\right.
\\
\nonumber
&\quad 
\left. -\left[(d_1 - 2 d_1^\prime) C_{\ell uud}^{\tR\tL,x} + (d_2-d_2^\prime) C_{\ell uud}^{\tR\tR,x} \right] (\overline{\ell_{\tR x}^\C} \gamma^\mu p_\tL) \right\} \phi_\mu^{(8)}
\\
\nonumber
&\quad + (d_1 C_{\nu dud}^{\tL\tR,x} + d_2 C_{\nu dud}^{\tL\tL,x} ) (\overline{\nu_{\tL x}^\C} \gamma^\mu p_\tR) \rho^-_\mu 
+ (d_1 C^{\tL\tR,x}_{\nu sud} 
+ d_1^\prime C^{\tL\tR,x}_{\nu uds} +d_2 C^{\tL\tL,x}_{\nu sud} ) (\overline{\nu_{\tL x}^\C}\gamma^\mu p_R) K^{*-}_\mu
\\
\nonumber
&\quad + \left\{(d_1^\prime C^{\tL\tR,x}_{\ell usu} + d_2^\prime C^{\tL\tL,x}_{\ell usu} ) (\overline{\ell_{\tL x}^\C}\gamma^\mu p_{\tR}) - (d_1^\prime C^{\tR\tL,x}_{\ell usu} + d_2^\prime C^{\tR\tR,x}_{\ell usu} ) (\overline{\ell_{\tR x}^\C}\gamma^\mu p_{\tL}) \right\} \bar{K}_\mu^{*0}
\\
\nonumber
&\quad + \left\{ (d_1 C_{\ell uud}^{\tL\tR,x} + d_2 C_{\ell uud}^{\tL\tL,x} ) (\overline{\ell_{\tL x}^\C} \gamma^\mu n_{\tR}) - (d_1 C_{\ell uud}^{\tR\tL,x} + d_2 C_{\ell uud}^{\tR\tR,x} ) (\overline{\ell_{\tR x}^\C} \gamma^\mu n_{\tL})\right\} \rho_\mu^+
\\
\nonumber
&\quad -\frac{1}{\sqrt{2}} (d_1 C_{\nu dud}^{\tL\tR,x} +d_2 C_{\nu dud}^{\tL\tL,x} ) (\overline{\nu_{\tL x}^\C}\gamma^\mu n_\tR) \rho^0_\mu 
\\
\nonumber
&\quad + \frac{1}{\sqrt{6}} \left[ (d_1-2 d_1')C _{\nu dud}^{\tL\tR,x}  + (d_2-2 d_2') C _{\nu dud}^{\tL\tL,x}  \right](\overline{\nu_{\tL x}^\C} \gamma^\mu n_\tR )\phi_\mu^{(8)}
\\
&\quad + \left[ d_1 C^{\tL\tR,x}_{\nu sud} + d_1' C^{\tL\tR,x}_{\nu dsu} +d_2 C^{\tL\tL,x}_{\nu sud} +d_2' C^{\tL\tL,x}_{\nu dsu} \right](\overline{\nu_{\tL x}^\C} \gamma^\mu n_{\tR}) \bar{K}_\mu^{*0},
\\
\nonumber
{\cal L}_{{\tt N}lX}^{{\tt \Delta( B+L)=0}} & = 
-(d_1 C_{\bar\nu dud}^{\tR\tL,x} + d_2 C_{\bar\nu dud}^{\tR\tR,x} ) (\overline{\nu_{\tL x}} \gamma^\mu p_\tL) \rho^-_\mu 
-(d_1 C^{\tR\tL,x}_{\bar\nu sud} + d_1^\prime C^{\tR\tL,x}_{\bar\nu uds} +d_2 C^{\tR\tR,x}_{\bar\nu sud}) (\overline{\nu_{\tL x}} \gamma^\mu p_\tL) K^{*-}_\mu
\\
\nonumber
&\quad + \frac{1}{\sqrt{2}} (d_1 C_{\bar\nu dud}^{\tR\tL,x} +d_2 C_{\bar\nu dud}^{\tR\tR,x} ) (\overline{\nu_{\tL x}}\gamma^\mu n_\tL) \rho^0_\mu 
\\
\nonumber
&\quad - \frac{1}{\sqrt{6}} \left[ (d_1-2 d_1')C _{\bar\nu dud}^{\tR\tL,x}  + (d_2-2 d_2') C _{\bar\nu dud}^{\tR\tR,x}  \right](\overline{\nu_{\tL x}} \gamma^\mu n_\tL )\phi_\mu^{(8)}
\\
\nonumber
&\quad - \left[ d_1 C^{\tR\tL,x}_{\bar\nu sud} + d_1' C^{\tR\tL,x}_{\bar\nu dsu} + d_2 C^{\tR\tR,x}_{\bar\nu sud} +d_2' C^{\tR\tR,x}_{\bar\nu dsu} \right](\overline{\nu_{\tL x}} \gamma^\mu n_{\tL}) \bar{K}_\mu^{*0}
\\
&\quad + \left\{ (d_1' C_{\bar\ell dds}^{\tR\tL,x} + d_2' C_{\bar\ell dds}^{\tR\tR,x}) (\overline{\ell_{\tL x}}\gamma^\mu n_\tL) -  (d_1' C_{\bar\ell dds}^{\tL\tR,x} + d_2' C_{\bar\ell dds}^{\tL\tL,x}) (\overline{\ell_{\tR x}}\gamma^\mu n_\tR) \right\} K_\mu^{*-}.
\end{align} 
\end{subequations}
In practical calculation, we adopt the ideal mixing, $\phi_\mu^{(8)}=\omega_\mu/\sqrt{3}-\phi_\mu \sqrt{2}/\sqrt{3}$, to copy with the nucleon decay modes involving an $\omega$ meson. 
For the general expression of nucleon two-body decay involving a vector meson given in \cref{eq:GammaN2lX}, the coefficients appearing in each term of \cref{eq:CtNlX} can be extracted from the chiral Lagrangian terms above 
and are summarized in \cref{tab:B2lX_vertex} for each process.

\begin{table}[h]
\center
\resizebox{\linewidth}{!}{
\renewcommand{\arraystretch}{1.6}
\begin{tabular}{|c| c| c | c |}
\hline
$\texttt{N}\to lX$
& $C_{\texttt{N} \to BX}$
& \multicolumn{1}{c|}{$C^{\tL}_{Bl}$ (upper cell) and $C^{\tR}_{Bl}$ (lower cell)}
& \multicolumn{1}{c|}{$C^{\tL}_{\texttt{N}lX}$ (upper cell) and $C^{\tR}_{\texttt{N}lX} $ (lower cell)} 
\\\hline
\multirow{2}*{$p \to \ell_x^+ \rho^0$}  
& $C_{p \to p\rho^0}$
& $c_1 C^{\tL \tR,x}_{\ell uud} + 
  c_2C^{\tL \tL,x}_{\ell uud }$
& $-{1\over\sqrt{2}}(d_1 C^{\tR\tL,x}_{\ell uud} + d_2 C^{\tR\tR,x}_{\ell uud}  )$
\\\hhline{~---}
& ${1\over \sqrt{2}}{\textsc g}_{\textsc f}$
& $-\left( c_1 C^{\tR \tL,x}_{\ell uud} + 
  c_2C^{\tR \tR,x}_{\ell uud} \right)$
& $ {1\over\sqrt{2}} (d_1 C^{\tL\tR,x}_{\ell uud} + d_2 C^{\tL\tL,x}_{\ell uud}  )$
\\\hline
\multirow{2}*{$p \to \ell_x^+ \omega$}  
& $C_{p \to p\omega}$ 
& $c_1 C^{\tL \tR,x}_{\ell uud} + 
  c_2C^{\tL \tL,x}_{\ell uud }$
&  $- {1\over3\sqrt{2}} \left[(d_1 - 2 d_1^\prime) C_{\ell uud}^{\tR\tL,x} + (d_2- 2 d_2^\prime) C_{\ell uud}^{\tR\tR,x} \right]$
\\\hhline{~---}
& ${1\over \sqrt{2}}{\textsc g}_{\textsc f}$ 
& $-\left( c_1 C^{\tR \tL,x}_{\ell uud} + 
  c_2C^{\tR \tR,x}_{\ell uud} \right)$ 
& ${1\over 3\sqrt{2}} \left[(d_1 - 2 d_1^\prime) C_{\ell uud}^{\tL\tR,x} + (d_2- 2 d_2^\prime) C_{\ell uud}^{\tL\tL,x} \right]$
\\\hline
\multirow{2}*{$p \to e^+ K^{*0}$}  
& $C_{p  \to \Sigma^+ K^{*0}}$  
& $c_1 C^{\tL \tR,x}_{\ell usu} + 
  c_2C^{\tL \tL,x}_{\ell usu}$
& $-(d_1^\prime C^{\tR\tL,x}_{\ell usu} + d_2^\prime C^{\tR\tR,x}_{\ell usu} )$
\\\hhline{~---}
& $-{\textsc g}_{\textsc f}$
& $-\left(c_1 C^{\tR \tL,x}_{\ell usu} + 
  c_2C^{\tR \tR,x}_{\ell usu} \right)$
& $(d_1^\prime C^{\tL\tR,x}_{\ell usu} + d_2^\prime C^{\tL\tL,x}_{\ell usu} )$
\\\hline
\multirow{2}*{$\makecell{ p \to \bar\nu_x \rho^+ \\
\mbox{ (\colorbox{gray!15}{$p \to \nu_x \rho^+$})} }$}
& $C_{p \to n \rho^+}$ 
& $c_1 C^{\tL \tR,x}_{\nu dud} + 
  c_2C^{\tL \tL,x}_{\nu dud} $
& \cellcolor{gray!15} $-(d_1 C_{\bar\nu dud}^{\tR\tL,x} + d_2 C_{\bar\nu dud}^{\tR\tR,x} )$
\\\hhline{~---}
& ${\textsc g}_{\textsc f}$
&\cellcolor{gray!15}$-\left(c_1 C^{\tR \tL,x}_{\bar{\nu}dud} + 
  c_2 C^{\tR \tR,x}_{\bar{\nu}dud} \right)$
& $(d_1 C_{\nu dud}^{\tL\tR,x} + d_2 C_{\nu dud}^{\tL\tL,x} )$
\\\hline
\multirow{4}*{$\makecell{ p \to \bar\nu_x K^{*+}  \\
\mbox{~(\colorbox{gray!15}{$p \to \nu_x K^{*+} $})~} }$} 
& $C_{p  \to \Sigma^0 K^{*+}}$ 
& $\frac{1}{\sqrt{2}} \left[c_1 (C^{\tL \tR,x}_{\nu uds} - C^{\tL \tR,x}_{\nu dsu})
    - c_2 C^{\tL \tL,x}_{\nu dsu} \right]$  
& \cellcolor{gray!15}
\\\hhline{~--~}
&  $-{1\over \sqrt{2} }{\textsc g}_{\textsc f}$
&\cellcolor{gray!15} $-\frac{1}{\sqrt{2}} \left[c_1 (C^{\tR \tL,x}_{\bar{\nu}uds} - C^{\tR \tL,x}_{\bar{\nu}dsu}) - 
  c_2 C^{\tR \tR,x}_{\bar{\nu}dsu} \right]$ 
& \multirow{-2}*{\cellcolor{gray!15}$-(d_1 C^{\tR\tL,x}_{\bar\nu sud} + d_1^\prime C^{\tR\tL,x}_{\bar\nu uds} +d_2 C^{\tR\tR,x}_{\bar\nu sud})$}
\\\hhline{~---}
& $C_{p  \to \Lambda^0K^{*+}}$  
& $\frac{1}{\sqrt{6}} \left[c_1 
    (C^{\tL \tR,x}_{\nu uds} + C^{\tL \tR,x}_{\nu dsu} 
    - 2 C^{\tL \tR,x}_{\nu sud} )
    + c_2 ( C^{\tL \tL,x}_{\nu dsu}  - 2 C^{\tL \tL,x}_{\nu sud}) \right] $
& \multirow{2}*{$(d_1 C^{\tL\tR,x}_{\nu sud} 
+ d_1^\prime C^{\tL\tR,x}_{\nu uds} +d_2 C^{\tL\tL,x}_{\nu sud} )$}
\\\hhline{~--~}
&  $-\sqrt{3\over 2}{\textsc g}_{\textsc f}$
&\cellcolor{gray!15}$-\frac{1}{\sqrt{6}} \left[c_1 (C^{\tR \tL,x}_{\bar{\nu}uds} 
    +  C^{\tR \tL,x}_{\bar{\nu}dsu} - 2 C^{\tR \tL,x}_{\bar{\nu} sud} )+ 
  c_2 (C^{\tR \tR,x}_{\bar{\nu}dsu}  - 2 C^{\tR \tR,x}_{\bar{\nu} sud} ) \right]$ 
&
\\\hline
\multirow{2}*{$n \to \ell_x^+ \rho^{-}$} 
& $C_{n  \to p \rho^{-}}$  
& $c_1 C^{\tL \tR,x}_{\ell uud} + 
  c_2C^{\tL \tL,x}_{\ell uud }$
& $-(d_1 C_{\ell uud}^{\tR\tL,x} + d_2 C_{\ell uud}^{\tR\tR,x} )$
\\\hhline{~---}
&  ${\textsc g}_{\textsc f}$
& $-\left( c_1 C^{\tR \tL,x}_{\ell uud} + 
  c_2C^{\tR \tR,x}_{\ell uud} \right)$ 
& $(d_1 C_{\ell uud}^{\tL\tR,x} + d_2 C_{\ell uud}^{\tL\tL,x} )$
\\\hline
\multirow{2}*{$\makecell{ n \to \bar\nu_x \rho^{0}  \\
\mbox{ (\colorbox{gray!15}{$n \to \nu_x \rho^{0}$})} }$} 
& $C_{n  \to n \rho^{0}}$   
& $c_1 C^{\tL \tR,x}_{\nu dud} + 
  c_2C^{\tL \tL,x}_{\nu dud} $
& \cellcolor{gray!15}{${1\over\sqrt{2}} (d_1 C_{\bar\nu dud}^{\tR\tL,x} +d_2 C_{\bar\nu dud}^{\tR\tR,x} )$}
\\\hhline{~---}
& $-{1\over \sqrt{2}}{\textsc g}_{\textsc f}$
& \cellcolor{gray!15}$-\left(c_1 C^{\tR \tL,x}_{\bar{\nu}dud} + 
  c_2 C^{\tR \tR,x}_{\bar{\nu}dud} \right)$ 
&  $- {1\over\sqrt{2}} (d_1 C_{\nu dud}^{\tL\tR,x} +d_2 C_{\nu dud}^{\tL\tL,x} )$
\\\hline
\multirow{2}*{$\makecell{ n \to \bar\nu_x \omega  \\
\mbox{ (\colorbox{gray!15}{$n \to \nu_x \omega$})} }$} 
& $C_{n  \to n \omega}$     
& $c_1 C^{\tL \tR,x}_{\nu dud} + 
  c_2C^{\tL \tL,x}_{\nu dud} $
& \cellcolor{gray!15}{ $-{1\over 3\sqrt{2}}\left[ (d_1-2 d_1')C _{\bar\nu dud}^{\tR\tL,x}  + (d_2-2 d_2') C _{\bar\nu dud}^{\tR\tR,x}  \right]$ } 
\\\hhline{~---}
& ${1\over \sqrt{2}}{\textsc g}_{\textsc f}$ 
&\cellcolor{gray!15}$-\left(c_1 C^{\tR \tL,x}_{\bar{\nu}dud} + 
  c_2 C^{\tR \tR,x}_{\bar{\nu}dud} \right)$
  
&  ${1\over 3\sqrt{2}}\left[ (d_1-2 d_1')C _{\nu dud}^{\tL\tR,x}  + (d_2-2 d_2') C _{\nu dud}^{\tL\tL,x}  \right]$
\\\hline
\multirow{4}*{$\makecell{ n \to \bar\nu_x K^{*0}  \\
\mbox{ (\colorbox{gray!15}{$n \to \nu_x K^{*0}$})} }$} 
& $C_{n  \to \Sigma^0 K^{*0}}$  
& $\frac{1}{\sqrt{2}} \left[c_1 (C^{\tL \tR,x}_{\nu uds} - C^{\tL \tR,x}_{\nu dsu})
    - c_2 C^{\tL \tL,x}_{\nu dsu} \right]$  
& \cellcolor{gray!15}
\\\hhline{~--~}
& ${1\over \sqrt{2} }{\textsc g}_{\textsc f}$ 
&\cellcolor{gray!15} $~-\frac{1}{\sqrt{2}} \left[c_1 (C^{\tR \tL,x}_{\bar{\nu}uds} 
- C^{\tR \tL,x}_{\bar{\nu}dsu}) 
- c_2 C^{\tR \tR,x}_{\bar{\nu}dsu} \right]~$   
& \multirow{-2}*{\cellcolor{gray!15}$~-\left[ d_1 C^{\tR\tL,x}_{\bar\nu sud} + d_1' C^{\tR\tL,x}_{\bar\nu dsu} + d_2 C^{\tR\tR,x}_{\bar\nu sud} +d_2' C^{\tR\tR,x}_{\bar\nu dsu} \right]~$ }
\\\hhline{~---}
& $C_{n  \to \Lambda^0 K^{*0}}$
& $\frac{1}{\sqrt{6}} \left[c_1 
    (C^{\tL \tR,x}_{\nu uds} + C^{\tL \tR,x}_{\nu dsu} 
    - 2 C^{\tL \tR,x}_{\nu sud} )
    + c_2 ( C^{\tL \tL,x}_{\nu dsu}  - 2 C^{\tL \tL,x}_{\nu sud}) \right] $ 
& 
\\\hhline{~--~}
& $-\sqrt{3\over 2}{\textsc g}_{\textsc f}$  
&\cellcolor{gray!15}$-\frac{1}{\sqrt{6}} \left[c_1 (C^{\tR \tL,x}_{\bar{\nu}uds} 
    +  C^{\tR \tL,x}_{\bar{\nu}dsu} - 2 C^{\tR \tL,x}_{\bar{\nu} sud} )+ 
  c_2 (C^{\tR \tR,x}_{\bar{\nu}dsu}  - 2 C^{\tR \tR,x}_{\bar{\nu} sud} ) \right]$  
& \multirow{-2}*{$\left[ d_1 C^{\tL\tR,x}_{\nu sud} + d_1' C^{\tL\tR,x}_{\nu dsu} +d_2 C^{\tL\tL,x}_{\nu sud} +d_2' C^{\tL\tL,x}_{\nu dsu} \right]$}
\\\hline
\cellcolor{gray!15}  
&\cellcolor{gray!15}$~C_{n  \to \Sigma^- K^{*+}}~$  
&\cellcolor{gray!15}$c_1 C^{\tL \tR,x}_{\bar{\ell} dds} + 
  c_2 C^{\tL \tL,x}_{\bar{\ell} dds}$
& \cellcolor{gray!15}$-(d_1' C_{\bar\ell dds}^{\tR\tL,x} + d_2' C_{\bar\ell dds}^{\tR\tR,x})$
\\\hhline{~---}
\multirow{-2}*{\cellcolor{gray!15}$n \to e^- K^{*+}$}   
&\cellcolor{gray!15} $-{\textsc g}_{\textsc f}$ 
&\cellcolor{gray!15} $-\left(c_1 C^{\tR \tL,x}_{\bar{\ell} dds} + 
  c_2 C^{\tR \tR,x}_{\bar{\ell} dds} \right) $
& \cellcolor{gray!15}$(d_1' C_{\bar\ell dds}^{\tL\tR,x} + d_2' C_{\bar\ell dds}^{\tL\tL,x})$
\\\hline
\end{tabular}
}
\caption{The explicit expression for the relevant vertices in the two-body nucleon decay involving a vector meson. The white and gray entries correspond to the processes with $\Delta(B-L)=0$ and $\Delta(B+L)=0$, respectively.
From the second column, one can recognize the intermediate baryon contributing to the diagram (a) in \cref{fig:N2lX_diagram}.}
\label{tab:B2lX_vertex}
\end{table}

For completeness, we also list the BNV three-point interactions involving a nucleon field and a pseudoscalar field, which are relevant to the two-body decays $\texttt{N}\rightarrow l M$. By expanding \cref{eq:LBlM} to the linear order in the pseudoscalar meson field, we have
\begin{subequations}
\begin{align}
\nonumber
{\cal L}_{\texttt{N}lM}^{{\tt \Delta( B-L)=0}} &= {i \over F_0}
\bigg\{ \frac{1}{2} \left[(c_1 C^{\tL\tR,x}_{\ell uud} + c_2 C^{\tL\tL,x}_{\ell uud}) (\overline{\ell_{\tL x}^\C} p_{\tL}) + (c_1 C^{\tR\tL,x}_{\ell uud} + c_2 C^{\tR\tR,x}_{\ell uud}) (\overline{\ell_{\tR x}^\C} p_{\tR})\right]\pi^0 
\\
\nonumber
&\quad - \frac{1}{2\sqrt{3}} \left[(c_1 C^{\tL\tR,x}_{\ell uud}-3 c_2 C^{\tL\tL,x}_{\ell uud}) (\overline{\ell_{\tL x}^\C} p_{\tL}) + (c_1 C^{\tR\tL,x}_{\ell uud}-3 c_2 C^{\tR\tR,x}_{\ell uud}) (\overline{\ell_{\tR x}^\C} p_{\tR}) \right]\eta
\\
\nonumber
&\quad + \frac{1}{\sqrt{2}} \left[ (c_1 C^{\tL\tR,x}_{\ell usu} -c_2 C^{\tL\tL,x}_{\ell usu}) (\overline{\ell_{\tL x}^\C} p_{\tL}) +  (c_1 C^{\tR\tL,x}_{\ell usu} -c_2 C^{\tR\tR,x}_{\ell usu}) (\overline{\ell_{\tR x}^\C} p_{\tR})\right] \bar{K}^0 
\\
\nonumber
&\quad + \frac{1}{\sqrt{2}} (c_1 C^{\tL\tR,x}_{\nu dud}+c_2 C^{\tL\tL,x}_{\nu dud}) (\overline{\nu_{\tL x}^\C} p_{\tL})\pi^- 
+ {1 \over \sqrt{2}} \left[c_1 (C^{\tL\tR,x}_{\nu uds} + C^{\tL\tR,x}_{\nu sud}) + c_2 C^{\tL\tL,x}_{\nu sud} \right] (\overline{\nu_{\tL x}^\C} p_{\tL}) K^- \bigg\}
\\
\nonumber
&\quad + {i \over F_0}
\bigg\{ - \frac{1}{2} (c_1 C^{\tL\tR,x}_{\nu dud} + c_2 C^{\tL\tL,x}_{\nu dud}) (\overline{\nu_{\tL x}^\C} n_{\tL})\pi^0 
- \frac{1}{2\sqrt{3}} (c_1 C^{\tL\tR,x}_{\nu dud} - 3 c_2 C^{\tL\tL,x}_{\nu dud} ) (\overline{\nu_{\tL x}^\C} n_{\tL}) \eta 
\\
\nonumber
&\quad + \frac{1}{\sqrt{2}} \left[ (c_1 C^{\tL\tR,x}_{\ell uud} +c_2 C^{\tL\tL,x}_{\ell uud}) (\overline{\ell_{\tL x}^\C} n_{\tL}) +  (c_1 C^{\tR\tL,x}_{\ell uud} +c_2 C^{\tR\tR,x}_{\ell uud}) (\overline{\ell_{\tR x}^\C} n_{\tR})\right]\pi^+
\\
&\quad + \frac{1}{\sqrt{2}} \left[ c_1( C^{\tL\tR,x}_{\nu sud}  + C^{\tL\tR,x}_{\nu dsu}) + c_2 (C^{\tL\tL,x}_{\nu sud} - C^{\tL\tL,x}_{\nu dsu}) \right] (\overline{\nu_{\tL x}^\C} n_{\tL}) \bar{K}^0 \bigg\} ,
\\
\nonumber
{\cal L}_{\texttt{N}lM}^{{\tt \Delta( B+L)=0}} &= 
{i \over F_0}\bigg\{ \frac{1}{\sqrt{2}} (c_1 C^{\tR\tL,x}_{\bar\nu dud}+c_2 C^{\tR\tR,x}_{\bar\nu dud}) (\overline{\nu_{\tL x}} p_{\tR})\pi^- 
+ \frac{1}{\sqrt{2}} [ c_1 ( C^{\tR\tL,x}_{\bar\nu uds} + C^{\tR\tL,x}_{\bar\nu sud} ) + c_2 C^{\tR\tR,x}_{\bar\nu sud} ] (\overline{\nu_{\tL x}} p_{\tR}) K^- \bigg\}
\\
\nonumber
&\quad + \frac{i}{F_0} \bigg\{ -{ 1 \over 2} (c_1 C^{\tR\tL,x}_{\bar\nu dud} + c_2 C^{\tR\tR,x}_{\bar\nu dud}) (\overline{\nu_{\tL x}} n_{\tR})\pi^0  
- {1 \over 2\sqrt{3}} (c_1 C^{\tR\tL,x}_{\bar\nu dud} - 3 c_2 C^{\tR\tR,x}_{\bar\nu dud} ) (\overline{\nu_{\tL x}} n_{\tR}) \eta 
\\
\nonumber
&\quad +\frac{1}{\sqrt{2}} \left[ c_1( C^{\tR\tL,x}_{\bar\nu sud}  + C^{\tR\tL,x}_{\bar\nu dsu}) + c_2 (C^{\tR\tR,x}_{\bar\nu sud} - C^{\tR\tR,x}_{\bar\nu dsu}) \right] (\overline{\nu_{\tL x}} n_{\tR}) \bar{K}^0
\\
&\quad +
\frac{1}{\sqrt{2} } \left[ (c_1 C^{\tL\tR,x}_{\bar\ell dds} -c_2 C^{\tL\tL,x}_{\bar\ell dds}) (\overline{\ell_{\tR x}} n_{\tL})  +  (c_1 C^{\tR\tL,x}_{\bar\ell dds} -c_2 C^{\tR\tR,x}_{\bar\ell dds}) (\overline{\ell_{\tL x}} n_{\tR}) \right] K^-\bigg\}.
\end{align}
\end{subequations}

\begin{table}[t]
\center
\resizebox{\linewidth}{!}{
\renewcommand{\arraystretch}{1.6}
\begin{tabular}{|c| c| c | c |}
\hline
$\texttt{N}\to lM$
& $C_{\texttt{N} \to BM}$
& \multicolumn{1}{c|}{$C^{\tL}_{Bl}$ (upper cell) and $C^{\tR}_{Bl}$ (lower cell)}
& \multicolumn{1}{c|}{$C^{\tL}_{\texttt{N}lM}$ (upper cell) and $C^{\tR}_{\texttt{N}lM}$ (lower cell)} 
\\\hline
\multirow{2}*{$p\to \ell^+ \pi^0$ } 

& $C_{p\to p\pi^0}$
& $c_1 C^{\tL \tR,x}_{\ell uud} + 
  c_2C^{\tL \tL,x}_{\ell uud }$
& ${1 \over 2} (c_1 C^{\tL\tR,x}_{\ell uud} + c_2 C^{\tL\tL,x}_{\ell uud})$
\\\hhline{~---}
& ${D+F\over 2}$
& $-( c_1 C^{\tR \tL,x}_{\ell uud} + 
  c_2C^{\tR \tR,x}_{\ell uud} )$
& ${1 \over 2} (c_1 C^{\tR\tL,x}_{\ell uud} + c_2 C^{\tR\tR,x}_{\ell uud})$
\\\hline 
\multirow{2}*{$p\to \ell^+ \eta$}
& $C_{p\to p\eta}$
& $c_1 C^{\tL \tR,x}_{\ell uud} + 
  c_2C^{\tL \tL,x}_{\ell uud }$
& $-{1 \over 2\sqrt{3}} (c_1 C^{\tL\tR,x}_{\ell uud}-3 c_2 C^{\tL\tL,x}_{\ell uud})$
\\\hhline{~---}
& $-{D-3F\over 2\sqrt{3}}$
& $-( c_1 C^{\tR \tL,x}_{\ell uud} + 
  c_2C^{\tR \tR,x}_{\ell uud} )$
& $-{1 \over 2\sqrt{3}} (c_1 C^{\tR\tL,x}_{\ell uud}-3 c_2 C^{\tR\tR,x}_{\ell uud})$
\\\hline 
\multirow{2}*{$p\to \ell^+ K^0$}
& $C_{p\to \Sigma^+ K^0}$
& $c_1 C^{\tL \tR,x}_{\ell usu} + 
  c_2C^{\tL \tL,x}_{\ell usu}$
& ${1 \over \sqrt{2}} (c_1 C^{\tL\tR,x}_{\ell usu} -c_2 C^{\tL\tL,x}_{\ell usu})$
\\\hhline{~---}
& ${D-F\over \sqrt{2}}$
& $-(c_1 C^{\tR \tL,x}_{\ell usu} + 
  c_2C^{\tR \tR,x}_{\ell usu} )$
& ${1 \over \sqrt{2}} (c_1 C^{\tR\tL,x}_{\ell usu} -c_2 C^{\tR\tR,x}_{\ell usu})$
\\\hline 
\multirow{2}*{$\makecell{ p \to \bar\nu_x \pi^+ \\
\mbox{ (\colorbox{gray!15}{$p \to \nu_x \pi^+$})} }$}
& $C_{p\to n\pi^+}$
& $c_1 C^{\tL \tR,x}_{\nu dud} + 
  c_2C^{\tL \tL,x}_{\nu dud} $
& ${1 \over \sqrt{2}} (c_1 C^{\tL\tR,x}_{\nu dud}+c_2 C^{\tL\tL,x}_{\nu dud})$
\\\hhline{~---}
& ${D+F\over \sqrt{2}}$
& \cellcolor{gray!15}$-(c_1 C^{\tR \tL,x}_{\bar{\nu}dud} + 
  c_2 C^{\tR \tR,x}_{\bar{\nu}dud})$
& \cellcolor{gray!15}${1\over \sqrt{2}} (c_1 C_{\bar\nu dud}^{\tR\tL,x} + c_2 C_{\bar\nu dud}^{\tR\tR,x} )$
\\\hline 
\multirow{4}*{$\makecell{ p \to \bar\nu_x K^+ \\
\mbox{~(\colorbox{gray!15}{$p \to \nu_x K^+$})~} }$}
& $C_{p\to \Lambda^0 K^+}$
& $\frac{1}{\sqrt{6}} \left[c_1 
    (C^{\tL \tR,x}_{\nu uds} + C^{\tL \tR,x}_{\nu dsu} 
    - 2 C^{\tL \tR,x}_{\nu sud} )
    + c_2 ( C^{\tL \tL,x}_{\nu dsu}  - 2 C^{\tL \tL,x}_{\nu sud}) \right]$
&  \multirow{2}*{${1 \over \sqrt{2}}\left[  c_1 (C^{\tL\tR,x}_{\nu uds} + C^{\tL\tR,x}_{\nu sud}) + c_2 C^{\tL\tL,x}_{\nu sud} \right]$}
\\\hhline{~--}
& $-{D+3F\over 2\sqrt{3}}$
& \cellcolor{gray!15}$-\frac{1}{\sqrt{6}} \left[c_1 (C^{\tR \tL,x}_{\bar{\nu}uds} 
    +  C^{\tR \tL,x}_{\bar{\nu}dsu} - 2 C^{\tR \tL,x}_{\bar{\nu} sud} )+ 
  c_2 (C^{\tR \tR,x}_{\bar{\nu}dsu}  - 2 C^{\tR \tR,x}_{\bar{\nu} sud} ) \right]$
& 
\\\hhline{~---}
& $C_{p\to \Sigma^0 K^+}$
& $\frac{1}{\sqrt{2}} \left[c_1 (C^{\tL \tR,x}_{\nu uds} - C^{\tL \tR,x}_{\nu dsu})
    - c_2 C^{\tL \tL,x}_{\nu dsu} \right]$
& \cellcolor{gray!15}
\\\hhline{~--}
& ${D-F\over 2}$
& \cellcolor{gray!15}$-\frac{1}{\sqrt{2}} \left[c_1 (C^{\tR \tL,x}_{\bar{\nu}uds} - C^{\tR \tL,x}_{\bar{\nu}dsu}) - 
  c_2 C^{\tR \tR,x}_{\bar{\nu}dsu} \right]$
& \multirow{-2}*{\cellcolor{gray!15}{${1\over \sqrt{2}}\left[ c_1 ( C^{\tR\tL,x}_{\bar\nu uds} + C^{\tR\tL,x}_{\bar\nu sud} ) + c_2 C^{\tR\tR,x}_{\bar\nu sud}\right]$}}
\\\hline 
\multirow{2}*{$n\to \ell_x^+ \pi^-$}
& $C_{n\to p \pi^-}$
& $c_1 C^{\tL \tR,x}_{\ell uud} + 
  c_2C^{\tL \tL,x}_{\ell uud }$
&  ${1\over \sqrt{2}} (c_1 C^{\tL\tR,x}_{\ell uud} +c_2 C^{\tL\tL,x}_{\ell uud})$
\\\hhline{~---}
& ${D+F\over \sqrt{2}}$
& $-( c_1 C^{\tR \tL,x}_{\ell uud} + 
  c_2C^{\tR \tR,x}_{\ell uud} )$
& ${1\over \sqrt{2}}(c_1 C^{\tR\tL,x}_{\ell uud} +c_2 C^{\tR\tR,x}_{\ell uud})$
\\\hline 
\multirow{2}*{$\makecell{ n \to \bar\nu_x \pi^0 \\
\mbox{ (\colorbox{gray!15}{$n \to \nu_x \pi^0$})} }$}
& $C_{n\to n \pi^0}$
& $c_1 C^{\tL \tR,x}_{\nu dud} + 
  c_2C^{\tL \tL,x}_{\nu dud} $
& $ -{ 1 \over 2} (c_1 C^{\tL\tR,x}_{\nu dud} + c_2 C^{\tL\tL,x}_{\nu dud})$
\\\hhline{~---}
& $-{D+F\over 2}$
& \cellcolor{gray!15}$-(c_1 C^{\tR \tL,x}_{\bar{\nu}dud} + c_2 C^{\tR \tR,x}_{\bar{\nu}dud})$
& \cellcolor{gray!15}{$-{ 1 \over 2} (c_1 C^{\tR\tL,x}_{\bar\nu dud} + c_2 C^{\tR\tR,x}_{\bar\nu dud})$}
\\\hline 
\multirow{2}*{$\makecell{ n \to \bar\nu_x \eta \\
\mbox{ (\colorbox{gray!15}{$n \to \nu_x \eta$})} }$}
&  $C_{n\to n \eta}$
&  $c_1 C^{\tL \tR,x}_{\nu dud} + 
  c_2C^{\tL \tL,x}_{\nu dud} $
& $- {1\over 2\sqrt{3}} (c_1 C^{\tL\tR,x}_{\nu dud} - 3 c_2 C^{\tL\tL,x}_{\nu dud} )$
\\\hhline{~---}
& $-{D-3F\over 2\sqrt{3}}$
& \cellcolor{gray!15}$-(c_1 C^{\tR \tL,x}_{\bar{\nu}dud} + c_2 C^{\tR \tR,x}_{\bar{\nu}dud})$
& \cellcolor{gray!15}{$- {1 \over 2\sqrt{3}} (c_1 C^{\tR\tL,x}_{\bar\nu dud} - 3 c_2 C^{\tR\tR,x}_{\bar\nu dud} )$}
\\\hline 
\multirow{4}*{$\makecell{ n \to \bar\nu_x K^0 \\
\mbox{ (\colorbox{gray!15}{$n \to \nu_x K^0$})} }$}
& $C_{n\to \Lambda^0 K^0}$
& $\frac{1}{\sqrt{6}} \left[c_1 (C^{\tL \tR,x}_{\nu uds} + C^{\tL \tR,x}_{\nu dsu} - 2 C^{\tL \tR,x}_{\nu sud} )
+ c_2 ( C^{\tL \tL,x}_{\nu dsu}  - 2 C^{\tL \tL,x}_{\nu sud}) \right]$
& \multirow{2}*{$~{1\over \sqrt{2}} \left[ c_1( C^{\tL\tR,x}_{\nu sud}  + C^{\tL\tR,x}_{\nu dsu}) + c_2 (C^{\tL\tL,x}_{\nu sud} - C^{\tL\tL,x}_{\nu dsu}) \right]~$}
\\\hhline{~--}
& $-{D+3F\over 2\sqrt{3}}$
& \cellcolor{gray!15}$~-\frac{1}{\sqrt{6}} \left[c_1 (C^{\tR \tL,x}_{\bar{\nu}uds} +  C^{\tR \tL,x}_{\bar{\nu}dsu} - 2 C^{\tR \tL,x}_{\bar{\nu} sud} )+ c_2 (C^{\tR \tR,x}_{\bar{\nu}dsu}  - 2 C^{\tR \tR,x}_{\bar{\nu} sud} ) \right]~$
& 
\\\hhline{~---}
& $C_{n\to \Sigma^0 K^0}$
& $\frac{1}{\sqrt{2}} \left[c_1 (C^{\tL \tR,x}_{\nu uds} - C^{\tL \tR,x}_{\nu dsu}) - c_2 C^{\tL \tL,x}_{\nu dsu} \right]$
& \cellcolor{gray!15}
\\\hhline{~--}
& $-{D-F\over 2}$
& \cellcolor{gray!15}$-\frac{1}{\sqrt{2}} \left[c_1 (C^{\tR \tL,x}_{\bar{\nu}uds} - C^{\tR \tL,x}_{\bar{\nu}dsu}) - 
  c_2 C^{\tR \tR,x}_{\bar{\nu}dsu} \right]$
& \multirow{-2}*{\cellcolor{gray!15}{${1\over \sqrt{2}} \left[ c_1( C^{\tR\tL,x}_{\bar\nu sud}  + C^{\tR\tL,x}_{\bar\nu dsu}) + c_2 (C^{\tR\tR,x}_{\bar\nu sud} - C^{\tR\tR,x}_{\bar\nu dsu}) \right]$}}
\\\hline 
\cellcolor{gray!15}
& $~C_{n\to \Sigma^- K^+}~$
& $c_1 C^{\tL \tR,x}_{\bar{\ell} dds} + 
  c_2 C^{\tL \tL,x}_{\bar{\ell} dds}$
& \cellcolor{gray!15}{${1 \over \sqrt{2}} (c_1 C^{\tL\tR,x}_{\bar\ell dds} -c_2 C^{\tL\tL,x}_{\bar\ell dds})$}
\\\hhline{~---}
\multirow{-2}*{\cellcolor{gray!15}{$n\to \ell_x^- K^+$}}
& ${D-F\over \sqrt{2}}$
& $-(c_1 C^{\tR \tL,x}_{\bar{\ell} dds} + 
  c_2 C^{\tR \tR,x}_{\bar{\ell} dds} )$
& \cellcolor{gray!15}{${1 \over \sqrt{2}} (c_1 C^{\tR\tL,x}_{\bar\ell dds} -c_2 C^{\tR\tR,x}_{\bar\ell dds})$}
\\\hline
\end{tabular} }
\caption{Same as \cref{tab:B2lX_vertex} but for the decay modes involving a pseudoscalar meson.}
\label{tab:B2lM_vertex}
\end{table}

\section{General expression for nucleon two-body decay involving a pseudoscalar meson}
\label{app:N2lM}

For the two-body modes involving a pseudoscalar, the diagrams are the same as those shown in \cref{fig:N2lX_diagram}, with the vector meson replaced by a pseudoscalar meson. The effective vertices used for the calculation are parametrized as, 
\begin{align}
 {\cal L}_{{\tt N} \to l M }   
& = 
{ C_{\texttt{N}\to BM}\over F_0 }
 \overline{B}\gamma_\mu \gamma_5 \texttt{N} \, \partial^\mu \bar M
+ C_{Bl}^\tL \overline{l}B_\tL+ C_{Bl}^\tR \overline{l}B_\tR
+ {i\over F_0} C_{{\tt N}lM}^\tL \bar M \overline{l} {\tt N}_\tL
+ {i\over F_0} C_{{\tt N}lM}^\tR \bar M \overline{l} {\tt N}_\tR,
\label{eq:LB2lM}
\end{align}
where the coefficients are derived from the same chiral Lagrangians given in \cref{eq:LBlM,eq:LBlX,eq:LB0}, and are summarized in \cref{tab:B2lM_vertex} for each process, based on the dim-6 LEFT BNV interactions. By calculating the two diagrams analogous to those in \cref{fig:N2lX_diagram} for the process ${\tt N}\to lM$, we obtain the following general expression for the decay width, 
\begin{align}
\label{eq:GammaN2lM}
\Gamma_{{\tt N} \to l M} &= \frac{m_{\tt N} \lambda^{1/2}(1,x_l, x_M)} {16\pi }
\frac{ 1}{2F_0^2} 
\Big[ (1 + x_l -x_M)
\big( |\tilde C_{{\tt N}lM}^\tL|^2 
+ |\tilde C_{{\tt N}lM}^\tR|^2\big) 
+ 4 x_l^{1\over2}\Re(\tilde C_{{\tt N}lM}^\tL \tilde C_{{\tt N}lM}^{\tR,*}) \Big],
\end{align}
where $\lambda(x,y,z)\equiv x^2+y^2+z^2-2xy-2yz-2zx$ 
is the triangle function, $x_M= m_M^2/m_{\tt N}^2$, and $\tilde C_{{\tt N}lM}^{\tL,\tR} $ are related to the parameters in \cref{eq:LB2lM} via,
\begin{subequations}
\begin{align}
\tilde C_{{\tt N}lM}^\tL & 
=   C_{{\tt N}lM}^\tL  
+ \sum_B \frac{C_{{\tt N}\to BM}}{m_B^2 - m_l^2} 
\left[ (m_{\tt N} m_B + m_l^2) C_{Bl}^\tL + m_l(m_{\tt N} +m_B) C_{Bl}^\tR  \right]  ,
\\
\tilde C_{{\tt N}lM}^\tR & 
= C_{{\tt N}lM}^\tR  
- \sum_B \frac{ C_{{\tt N}\to BM} }{m_B^2 - m_l^2} 
\left[ (m_{\tt N} m_B + m_l^2) C_{Bl}^\tR + m_l(m_{\tt N} +m_B) C_{Bl}^\tL  \right].
\end{align}
\end{subequations}

\begin{table}[t]
\center
\resizebox{\linewidth}{!}{
\renewcommand{\arraystretch}{1.4}
\begin{tabular}{|l|c|c|c|c|c|c|c|c|c|c|c|c|c|c|}
\hline
\multirow{3}*{\qquad~Mode}  
& \multirow{3}*{ $\makecell{ \Gamma^{-1}_{\rm exp.} \\(10^{30}\,\rm yr)}$}
&\multicolumn{13}{c|}{ Constraint on the effective scale $\Lambda_{\tt eff} \equiv |C_i^j|^{-1/2}\,(10^{15}\,\rm GeV) $ }
\\\cline{3-15}
&
&$~\calO_{\ell uud}^{\tL\tR,x}~$ 
&$~\calO_{\ell uud}^{\tL\tL,x}~$ 
&$~\calO_{\nu dud}^{\tL\tR,x}~$ 
&$~\calO_{\nu dud}^{\tL\tL,x}~$
&$~\calO_{\ell usu}^{\tL\tR,x}~$ 
&$~\calO_{\ell usu}^{\tL\tL,x}~$ 
&$~\calO_{\nu uds}^{\tL\tR,x}~$ 
&$~\calO_{\nu dsu}^{\tL\tR,x}~$ 
&$~\calO_{\nu dsu}^{\tL\tL,x}~$ 
&$~\calO_{\nu sud}^{\tL\tR,x}~$
&$~\calO_{\nu sud}^{\tL\tL,x}~$
&\cellcolor{gray!15}$~\calO_{\bar\ell dds}^{\tL\tR,x}~$
&\cellcolor{gray!15}$~\calO_{\bar\ell dds}^{\tL\tL,x}~$
\\\cline{3-15}
& 
&$~\calO_{\ell uud}^{\tR\tL,x}~$ 
&$~\calO_{\ell uud}^{\tR\tR,x}~$ 
&\cellcolor{gray!15}$\calO_{\bar\nu dud}^{\tR\tL,x}~$ 
&\cellcolor{gray!15}$\calO_{\bar\nu dud}^{\tR\tR,x}~$
&$~\calO_{\ell usu}^{\tR\tL,x}~$ 
&$~\calO_{\ell usu}^{\tR\tR,x}~$ 
&\cellcolor{gray!15}$~\calO_{\bar\nu uds}^{\tR\tL,x}~$ 
&\cellcolor{gray!15}$~\calO_{\bar\nu dsu}^{\tR\tL,x}~$ 
&\cellcolor{gray!15}$~\calO_{\bar\nu dsu}^{\tR\tR,x}~$ 
&\cellcolor{gray!15}$~\calO_{\bar\nu sud}^{\tR\tL,x}~$
&\cellcolor{gray!15}$~\calO_{\bar\nu sud}^{\tR\tR,x}~$
&\cellcolor{gray!15}$~\calO_{\bar\ell dds}^{\tR\tL,x}~$
&\cellcolor{gray!15}$~\calO_{\bar\ell dds}^{\tR\tR,x}~$
\\\hline
$~p\to e^+ \pi^0 $ & $~~2.4\times10^4~~$ 
&\cellcolor{gray!50}$4.01$ &\cellcolor{gray!50}$4.03$ & $-$ & $-$ & $-$ & $-$ & $-$ & $-$ & $-$ & $-$ & $-$ & $-$ & $-$
\\
$~p\to \mu^+ \pi^0 $ & $1.6\times10^4$ 
&\cellcolor{gray!50}$3.61$ &\cellcolor{gray!50}$3.62$ & $-$ & $-$ & $-$ & $-$ & $-$ & $-$ & $-$ & $-$ & $-$ & $-$ & $-$
\\
$~p\to e^+ \eta $ & $1.0\times10^4$ 
& $0.87$ & $2.59$ & $-$ & $-$ & $-$ & $-$ & $-$ & $-$ & $-$ & $-$ & $-$ & $-$ & $-$
\\
$~p\to \mu^+ \eta $ & $4.7\times10^3$ 
& $0.76$ & $2.13$ & $-$ & $-$ & $-$ & $-$ & $-$ & $-$ & $-$ & $-$ & $-$ & $-$ & $-$
\\
$~p\to e^+ K^{0} $ & $1.0\times10^3$ 
& $-$ & $-$ & $-$ &  $-$ &\cellcolor{gray!50}$1.39$ &\cellcolor{gray!50}$1.10$ & $-$ & $-$ & $-$ & $-$ & $-$ & $-$ & $-$
\\
$~p\to \mu^+ K^{0} $ & $4.5\times10^3$ 
& $-$ & $-$ & $-$ & $-$ &\cellcolor{gray!50}$2.01$ &\cellcolor{gray!50}$1.62$ & $-$ & $-$ & $-$ & $-$ & $-$ & $-$ & $-$
\\
$~p\to \bar\nu_x (\colorbox{gray!15}{$\nu_x$}) \pi^{+} $ & $390$ 
& $-$ & $-$ & $1.70$ & $1.71$ & $-$ & $-$ & $-$ & $-$ & $-$ & $-$ & $-$ & $-$ & $-$
\\
$~p\to \bar\nu_x (\colorbox{gray!15}{$\nu_x$}) K^{+}~$ & $5.9\times10^3$ 
& $-$ & $-$ & $-$ & $-$ & $-$ & $-$ &\cellcolor{gray!50}$1.78$ &\cellcolor{gray!50}$1.24$ &\cellcolor{gray!50}$1.25$ &\cellcolor{gray!50}$2.46$ &\cellcolor{gray!50}$2.47$ & $-$ & $-$
\\
\hline
$~n\to e^+ \pi^-$ & $5.3\times10^3$ 
& $3.27$ & $3.28$ & $-$ & $-$ & $-$ & $-$ & $-$ & $-$ & $-$ & $-$ & $-$ & $-$ & $-$
\\
$~n\to \mu^+ \pi^-$ & $3.5\times10^3$ 
& $2.93$ & $2.95$ & $-$ & $-$ & $-$ & $-$ & $-$ & $-$ & $-$ & $-$ & $-$ & $-$ & $-$
\\
$~n\to \bar\nu_x (\colorbox{gray!15}{$\nu_x$}) \pi^0 $ & $1.1\times10^3$ 
& $-$ & $-$ &\cellcolor{gray!50}$1.85$ &\cellcolor{gray!50}$1.86$ & $-$ & $-$ & $-$ & $-$ & $-$ & $-$ & $-$ & $-$ & $-$
\\
$~n\to \bar\nu_x (\colorbox{gray!15}{$\nu_x$}) \eta $ & 158 
& $-$ & $-$ & $0.31$ & $0.92$ & $-$ & $-$ & $-$ & $-$ & $-$ & $-$ & $-$ & $-$ & $-$
\\
$~n\to \bar\nu_x (\colorbox{gray!15}{$\nu_x$}) K^0 $ & 86 
& $-$ & $-$ & $-$ & $-$ & $-$ & $-$ & $0.43$ & $0.62$ & $0.74$ & $0.85$ & $0.86$ & $-$ & $-$
\\
\cellcolor{gray!15}$~n\to e^- K^+$ & 32 
& $-$ & $-$ & $-$ & $-$ & $-$ & $-$ & $-$ & $-$ & $-$ & $-$ & $-$ &\cellcolor{gray!50}$0.57$ & \cellcolor{gray!50}$0.49$
\\
\cellcolor{gray!15}$~n\to \mu^- K^+$ & 57 
& $-$ & $-$ & $-$ & $-$ & $-$ & $-$ & $-$ & $-$ & $-$ & $-$ & $-$ &\cellcolor{gray!50}$0.66$ &\cellcolor{gray!50}$0.57$
\\
\hline
\end{tabular} }
\caption{Bounds on the effective scale associated with each dim-6 BNV operator in the LEFT, derived from the two-body nucleon decays involving a pseudoscalar meson. The most stringent bound derived for each operator is highlighted in darker gray. Operators related by parity transformation $\tL\leftrightarrow \tR$ yield identical bounds and are therefore grouped together. For fermionic bilinears involving a neutrino field, the parity transformation maps the operators containing $\overline{\nu_\tL^\C}$ into those with $\overline{\nu_\tL}$, resulting in operators that carry opposite lepton number. Those operators and their corresponding processes with $\Delta (B+L)=0$ are highlighted in light gray. For operators involving a charged lepton, $\ell$, the superscript $x$ is either $e$ or $\mu$ depending on the specific process; while for those operators involving a neutrino, $x$ can represent any of the three flavors. 
} 
\label{tab:BoundonWC}
\end{table}

Based on the above formulas and the current experimental bounds on the two-body nucleon decay modes involving pseudoscalar mesons, we derive constraints on the relevant LEFT interactions under the assumption of single operator dominance. In \cref{tab:BoundonWC}, we present the resulting bounds on the effective scale associated with each relevant dim-6 BNV operator $\calO_i^j$, defined as the inverse square root of its WC, $\Lambda_{\tt eff}\equiv |C_i^j|^{-1/2}$. 
The second column lists the current experimental lower bounds on the inverse decay widths of all relevant modes. Due to the stringent experimental limits on these nucleon decay modes, the effective scale $\Lambda_{\tt eff}$ for most operators can be constrained to be above $\calO(10^{15})$ GeV. By selecting the most stringent bound obtained for each WC, primarily derived from proton decay channels, we further predict new constraints on nucleon decay modes involving vector mesons, as shown in \cref{tab:N2lXbound}.

\end{document}